\documentclass[superscriptaddress,showkeys,showpacs,preprintnumbers,floatfix,twocolumn]{revtex4-1}           
\RequirePackage{lineno}
\usepackage{epsfig}
\usepackage{graphicx}
\usepackage{pslatex}
\usepackage{color}
\usepackage{amssymb}
\usepackage{multirow}
\usepackage{xspace}

\usepackage{titlesec}
\usepackage{graphicx}
\usepackage{slashbox}
\usepackage{array}
\usepackage{multirow}
\usepackage{float}
\usepackage{color}
\usepackage{subfigure}
\usepackage{dirtytalk}
\usepackage{blindtext}
\newcolumntype{P}[1]{>{\centering\arraybackslash}p{#1}}
\newcolumntype{M}[1]{>{\centering\arraybackslash}m{#1}}
\DeclareGraphicsExtensions{.pdf,.jpeg,.png}
\usepackage{epstopdf}
\usepackage{mathtools}

\newcommand{\snn}{$\sqrt{s_{\mathrm{NN}}}~$}

\newcommand{\pt}{$p_{\mathrm{T}}~$}
\newcommand{\pta}{$p_{\mathrm{T}}$}

\begin{document}

\title{Pion, kaon, and (anti) proton production in U+U collisions at \snn = 193~GeV measured with the STAR detector}

\affiliation{American University of Cairo, New Cairo 11835, New Cairo, Egypt}
\affiliation{Texas A\&M University, College Station, Texas 77843}
\affiliation{Brookhaven National Laboratory, Upton, New York 11973}
\affiliation{Ohio State University, Columbus, Ohio 43210}
\affiliation{University of Kentucky, Lexington, Kentucky 40506-0055}
\affiliation{Joint Institute for Nuclear Research, Dubna 141 980}
\affiliation{Panjab University, Chandigarh 160014, India}
\affiliation{Variable Energy Cyclotron Centre, Kolkata 700064, India}
\affiliation{Alikhanov Institute for Theoretical and Experimental Physics NRC "Kurchatov Institute", Moscow 117218}
\affiliation{Abilene Christian University, Abilene, Texas   79699}
\affiliation{Instituto de Alta Investigaci\'on, Universidad de Tarapac\'a, Arica 1000000, Chile}
\affiliation{University of California, Riverside, California 92521}
\affiliation{University of Houston, Houston, Texas 77204}
\affiliation{University of Jammu, Jammu 180001, India}
\affiliation{State University of New York, Stony Brook, New York 11794}
\affiliation{National Research Nuclear University MEPhI, Moscow 115409}
\affiliation{Shanghai Institute of Applied Physics, Chinese Academy of Sciences, Shanghai 201800}
\affiliation{Yale University, New Haven, Connecticut 06520}
\affiliation{University of California, Davis, California 95616}
\affiliation{Lawrence Berkeley National Laboratory, Berkeley, California 94720}
\affiliation{University of California, Los Angeles, California 90095}
\affiliation{Indiana University, Bloomington, Indiana 47408}
\affiliation{Shandong University, Qingdao, Shandong 266237}
\affiliation{Fudan University, Shanghai, 200433 }
\affiliation{University of Science and Technology of China, Hefei, Anhui 230026}
\affiliation{Tsinghua University, Beijing 100084}
\affiliation{University of California, Berkeley, California 94720}
\affiliation{University of Heidelberg, Heidelberg 69120, Germany }
\affiliation{NRC "Kurchatov Institute", Institute of High Energy Physics, Protvino 142281}
\affiliation{Wayne State University, Detroit, Michigan 48201}
\affiliation{Indian Institute of Science Education and Research (IISER), Berhampur 760010 , India}
\affiliation{Kent State University, Kent, Ohio 44242}
\affiliation{Rice University, Houston, Texas 77251}
\affiliation{University of Tsukuba, Tsukuba, Ibaraki 305-8571, Japan}
\affiliation{University of Illinois at Chicago, Chicago, Illinois 60607}
\affiliation{Lehigh University, Bethlehem, Pennsylvania 18015}
\affiliation{University of Calabria \& INFN-Cosenza, Italy}
\affiliation{National Cheng Kung University, Tainan 70101 }
\affiliation{Purdue University, West Lafayette, Indiana 47907}
\affiliation{Southern Connecticut State University, New Haven, Connecticut 06515}
\affiliation{Central China Normal University, Wuhan, Hubei 430079 }
\affiliation{Temple University, Philadelphia, Pennsylvania 19122}
\affiliation{Valparaiso University, Valparaiso, Indiana 46383}
\affiliation{Indian Institute of Science Education and Research (IISER) Tirupati, Tirupati 517507, India}
\affiliation{Institute of Modern Physics, Chinese Academy of Sciences, Lanzhou, Gansu 730000 }
\affiliation{National Institute of Science Education and Research, HBNI, Jatni 752050, India}
\affiliation{Rutgers University, Piscataway, New Jersey 08854}
\affiliation{ELTE E\"otv\"os Lor\'and University, Budapest, Hungary H-1117}
\affiliation{University of Texas, Austin, Texas 78712}
\affiliation{Warsaw University of Technology, Warsaw 00-661, Poland}
\affiliation{Max-Planck-Institut f\"ur Physik, Munich 80805, Germany}
\affiliation{Creighton University, Omaha, Nebraska 68178}
\affiliation{Indian Institute Technology, Patna, Bihar 801106, India}
\affiliation{Ball State University, Muncie, Indiana, 47306}
\affiliation{Huzhou University, Huzhou, Zhejiang  313000}
\affiliation{Michigan State University, East Lansing, Michigan 48824}
\affiliation{Argonne National Laboratory, Argonne, Illinois 60439}
\affiliation{South China Normal University, Guangzhou, Guangdong 510631}
\affiliation{Frankfurt Institute for Advanced Studies FIAS, Frankfurt 60438, Germany}

\author{M.~S.~Abdallah}\affiliation{American University of Cairo, New Cairo 11835, New Cairo, Egypt}
\author{B.~E.~Aboona}\affiliation{Texas A\&M University, College Station, Texas 77843}
\author{J.~Adam}\affiliation{Brookhaven National Laboratory, Upton, New York 11973}
\author{J.~R.~Adams}\affiliation{Ohio State University, Columbus, Ohio 43210}
\author{J.~K.~Adkins}\affiliation{University of Kentucky, Lexington, Kentucky 40506-0055}
\author{G.~Agakishiev}\affiliation{Joint Institute for Nuclear Research, Dubna 141 980}
\author{I.~Aggarwal}\affiliation{Panjab University, Chandigarh 160014, India}
\author{M.~M.~Aggarwal}\affiliation{Panjab University, Chandigarh 160014, India}
\author{Z.~Ahammed}\affiliation{Variable Energy Cyclotron Centre, Kolkata 700064, India}
\author{A.~Aitbaev}\affiliation{Joint Institute for Nuclear Research, Dubna 141 980}
% ...

\collaboration{STAR Collaboration}\noaffiliation



\begin{abstract}
We present the first measurements of transverse momentum spectra of $\pi^{\pm}$, $K^{\pm}$, $p(\bar{p})$ at midrapidity ($|y| < 0.1$) in U+U collisions at \snn = 193~GeV with the STAR detector at the Relativistic Heavy Ion Collider (RHIC). The centrality dependence of particle yields, average transverse momenta, particle ratios and kinetic freeze-out parameters are discussed. The results are compared with the published results from Au+Au collisions at \snn~=~200~GeV in STAR. 
The results are also compared to those from A Multi Phase Transport (AMPT) model.
\end{abstract}

\maketitle

\section{INTRODUCTION}

The experimental need to study the medium formed in high-energy heavy-ion collisions 
is called for by theoretical predictions from Quantum Chromodynamics (QCD) ~\cite{Shuryak:1980tp,Gyulassy:2004zy,Blaizot:2001nr,Bazavov:2011nk,Aoki:2006we, Arsene:2004fa,Adcox:2004mh,Back:2004je,Adams:2005dq,Kharzeev:2000ph}. One of the main motivations lies in the exploration of the QCD phase diagram~\cite{Rajagopal:2000wf, Laermann:2003cv, Stephanov:2007fk}. The phase diagram in temperature ($T$) and baryon chemical potential ($\mu_{B}$) can be accessed experimentally by varying the collision energy of heavy-ion collisions~\cite{star-note1,star-note2,Abelev:2009bw,nature}. Earlier experimental results from the Relativistic Heavy Ion Collider (RHIC) confirmed the presence of a hot and dense deconfined medium dominated by partonic degrees of freedom~\cite{Gyulassy:2004zy,Arsene:2004fa,Adcox:2004mh,Back:2004je,Adams:2005dq}. The fluid of deconfined quarks and gluons, called the Quark Gluon Plasma (QGP), undergoes a transition to a gas of hadrons.
The measurements that evidence the formation of the QGP are: suppression of high transverse momentum (\pta) hadron production in nucleus-nucleus collisions relative to p+p collisions~\cite{Adler:2002xw,Adler:2002tq,Adams:2003kv,Adams:2003am,Abelev:2006jr,Abelev:2007ra,Adams:2006yt,Adler:2003qi,Arsene:2003yk,Back:2003qr,Aamodt:2010jd,CMS:2012aa}, the relatively large value of elliptic flow ($v_{2}$) associated with light quark as well as strange quark carrying hadrons, and the inequality between baryon and meson $v_{2}$ at intermediate \pta~\cite{Ackermann:2000tr,Adler:2001nb,Adler:2002pb,Adams:2003zg,Adams:2005zg,Abelev:2007rw,Adamczyk:2015ukd,Adamczyk:2017xur,Alver:2006wh,Adler:2003kt,Abelev:2014pua}.
 
A complementary approach to study heavy ion collisions at RHIC relies on varying the colliding system type, shape and geometry. The initial stage dynamics is expected to be reflected in the final state. Hence, the initial geometry and shape of the colliding nuclei could provide insights into the physics of the bulk properties of the system~\cite{Masui:2009qk,Haque:2011ti,Haque:2011aa,Bairathi:2015uba,Adamczyk:2016eux,Agakishiev:2011ar,Adare:2006ti,Adare:2008ad,Adare:2018toe,Adamczyk:2015obl, Adam:2018tdm}. With such a motivation~\cite{Heinz:2004ir,Kuhlman:2005ts,Kuhlman:2006qp,Nepali:2006ep,Nepali:2007an,Hirano:2010jg}, the Solenoidal Tracker At RHIC (STAR) took data in the year 2012 from uranium on uranium (U+U) collisions at \snn = 193~GeV
because Uranium nucleus ($_{92}\rm{U}^{238}$) is prolate in contrast to the nearly spherical Gold nucleus ($_{79}\rm{Au}^{197}$). 

The prolate shape of the uranium nucleus is associated with a long or major axis and a short or minor axis. Due to this, very interesting orientations of the two colliding nuclei are possible even for most central U+U collisions~\cite{Bairathi:2015uba}. Among all possible random orientations, three particular orientations are worth noting. Those are termed as : tip-tip, body-body and body-tip. Tip-tip refers to the configuration where the major axes of the two colliding nuclei lie parallel to the beam axis resulting in the tips of both nuclei colliding head-on. Body-body collisions are those with the major axes perpendicular to the beam axis and the minor axes parallel to the beam axis. The body-tip orientation of the two colliding nuclei is straightforward from the above two explanations. Model calculations suggest that a tip-tip collision leads to the production of the highest particle multiplicity~\cite{Haque:2011aa,Bairathi:2015uba}, and the maximum transverse particle density in U+U collisions could be $6\%-35\%$ higher than in Au+Au collisions~\cite{Heinz:2004ir,Nepali:2006ep,Nepali:2007an}. With all possible orientations combined, U+U collisions will still likely give higher energy density and particle multiplicity.

This paper concentrates on the extraction of the bulk properties of the medium in U+U collisions at \snn = 193~GeV, including all orientations of the colliding nuclei. The results presented here include the transverse momentum spectra, the particle yields ($dN/dy$), the mean transverse momentum ($\langle p_{\mathrm T}\rangle$), the particle ratios, and the kinetic freeze-out parameters. These results are obtained for $\pi^{\pm}$, $K^{\pm}$ and $p$($\bar{p}$) as a function of collision centrality. All the results are compared to the corresponding published results of Au+Au collisions at \snn = 200~GeV~\cite{Abelev:2008ab,Adams:2003xp}. A comparative study with AMPT model~\cite{Lin:2004en} modified to incorporate the deformation of uranium nucleus~\cite{Haque:2011aa} is also carried out. 

The paper is organized as follows. Section II discusses the STAR detector, event selection, centrality selection, and particle identification (PID)  procedures. Section III lists the details of the various correction factors to obtain the transverse momentum spectra of $\pi^{\pm}$, $K^{\pm}$ and $p$($\bar{p}$). The systematic uncertainties associated with the measurements of the yields are discussed in Sec. IV. The results in terms of \pt spectra, $dN/dy$, $\langle p_{\mathrm T}\rangle$, ratios of particle yields, kinetic freeze-out conditions and comparison to model calculations are presented in Sec. V. Finally in Sec. VI a summary is given.

\section{EXPERIMENT AND DATA ANALYSIS}
\subsection{The STAR Detector}

The results presented in this paper are from the analysis of data collected with the STAR detector at RHIC~\cite{Ackermann:2002ad}  for U+U collisions at \snn = 193~GeV in 2012.  The data are obtained with a minimum bias trigger using the Vertex Position Detectors (VPDs) ~\cite{Llope:2003ti,Llope:2014nva} and Zero Degree Calorimeters (ZDCs)~\cite{Adler:2000bd,Adler:2003sp}.
The two VPDs are located at 4.24 $<$ $|\eta|$ $<$ 5.1. The VPDs also provide the start time of the collision and the position of the collision vertex along the beam direction.
The ZDCs are a pair of hadronic calorimeters placed on either side of the beam pipe at a distance of 18 m from the center of the STAR detector that detect hadrons emerging at small angles ($\theta < 2$ mrad) with respect to the beam axis.  Particle tracking is done using the cylindrical Time Projection Chamber (TPC)~\cite{Anderson:2003ur} that is filled with P10 gas ($90\%$ argon and $10\%$ methane) and maintained at a pressure of 2~mbar above atmospheric pressure.  It functions in a uniform 0.5 T magnetic field that is parallel to the beam axis.  The TPC has full 2$\pi$ azimuthal coverage for a pseudorapidity range $|\eta|$ $<$ 1.0.  Particles are identified in the TPC by their ionization energy loss per unit track length ($\langle dE/dx \rangle$) along the particle trajectory.  The identification of particles at higher momentum was achieved by including the Time Of Flight (TOF) detector~\cite{Wu:2005ci,Llope:2003ti} that surrounds the TPC cylinder.  The TOF uses a Multigap Restive Plate Chamber (MRPC) technology and has pseudorapidity acceptance of $|\eta| < 0.9$ with full azimuthal coverage~\cite{Wu:2005ci,Llope:2003ti}.

\subsection{Event Selection}

The criteria for minimum bias triggered event selection begin with the identification of a primary vertex that is the common point of origin of tracks in an event.
Conventionally the $Z$-axis is taken to be along the beam axis with $z$ = 0 at the center of the TPC.  The $Z$ position of the primary vertex ($V_{z}$) was constrained to be within $|V_{z}|$ $< 30$ cm.
In addition, to reduce pile-up events, a cut of $|V_{z}-V_{z,VPD}| < 3$~cm is applied. Here $V_{z,VPD}$ is the vertex position along the beam axis measured by the VPD. The plane perpendicular to beam direction defines the transverse x-y direction. The radial vertex position $V_{r} = \sqrt{V_{x}^{2}+V_{y}^{2}}$ was required to be less than 2 cm to avoid including interactions between beam and beam pipe.
The number of minimum bias U+U collisions at \snn = 193~GeV analysed after these event selections is about 2.70 $\times$ $10^8$.

\subsection{Centrality Selection}

The collision centrality is qualitatively a measure of the overlap of the two colliding nuclei or, correspondingly, a measure of the number of colliding nucleons in the U+U collisions.  Quantitatively this is determined based on the measured uncorrected charged particle multiplicity in the TPC over the full azimuthal coverage and within $|\eta| < 0.5$. This is termed as Reference Multiplicity or simply Refmult in STAR.
The centrality selection is done by comparing the Refmult distribution with a Monte Carlo Glauber model simulation~\cite{Abelev:2009bw, Adamczyk:2015obl,Adamczyk:2016dzv}.
The MC Glauber model is based on charged particle multiplicity calculated using the two-component model with the number of participants ($N_{\text{part}}$) and number of binary nucleon-nucleon collisions ($N_{\text{coll}}$).
In this analysis the minimum bias events are divided into nine centrality classes with each class representing the fraction of the total collision cross section parametrized in terms of the number of participating nucleons in the collision, $\langle N_{\text{part}}\rangle$, with the class 0--5\% being the class with the highest particle multiplicities and therefore, the most central class.  The nine classes are 0--5\%, 5--10\%, 10--20\%, 20--30\%, 30--40\%, 40--50\%, 50--60\%, 60--70\% and 70--80\%.
The events for 70--80\% centrality class are corrected for the trigger inefficiencies and 80--100\% centrality is not used in our analysis because of its significant trigger bias due to vertex inefficiency at low multiplicities and the contamination from electromagnetic interactions.
Using the Glauber Model Monte-Carlo simulations, the average number of participant nucleons $\langle N_{\text{part}}\rangle$ and number of nucleon-nucleon binary collisions $\langle N_{\text{coll}}\rangle$ are evaluated. Systematic uncertainties on the average quantities have been estimated by varying input parameters for the Glauber model as well as the total cross section within $\pm$ 5\% and by using different density profiles for the nucleons in the uranium nuclei used in the simulations discussed in Ref.~\cite{Masui:2009qk}.  These values for  U+U collisions at \snn = 193~GeV along with the corresponding number of events analyzed in each centrality bin are listed in Table~\ref{tab:cent-event}.   The U+U collisions allow measurements to a higher value of $\langle N_{\text{part}}\rangle  > 400$ than in Au+Au collisions at \snn = 200~GeV ($\langle N_{\text{part}}\rangle \approx 350$). 

\begin{table}
\caption{The average number of participating nucleons $\langle N_{\rm{part}}\rangle$, average number of nucleon-nucleon binary collisions   $\langle N_{\text{coll}}\rangle$ and number of events in different collision centrality classes in U+U collisions at \snn = 193~GeV.}
 \centering   
\begin{tabular}{cccc}
    \hline	
Centrality (\%) & $\langle N_{\text{part}}\rangle $ & $\langle N_{\text{coll}}\rangle$ & Events ($\times 10^6$) \\
   \hline 
0--5 & 414.9 $\pm$ 6.0~~ &~1281.3~$\pm$~100 &  15.7  \\
5--10 & 355.4 $\pm$ 13.9 &~~1011.0~$\pm$~49.7&  17.8  \\
10--20 & 277.5 $\pm$ 12.8 &~~~~714.1 $\pm$~46.6 &  36.0  \\
20--30 & 195.4 $\pm$ 13.7 & ~~~~435.9~$\pm$~42.5 &  35.5  \\
30--40 & 133.1 $\pm$ 13.6 &~~~~253.5~$\pm$~36.3 &  35.8  \\
40--50 & ~~86.2 $\pm$ 12.6 & ~~~~137.4~$\pm$~28.6 &  35.1  \\
50--60 & ~~52.6 $\pm$ 10.5 & ~~~~~69.3~$\pm$~20.3 &  33.7  \\
60--70 & 29.4 $\pm$ 8.5 & ~~~~~31.8~$\pm$~11.8 &  33.4  \\
70--80 & 14.7 $\pm$ 5.5 & ~~~13.2~$\pm$~7.0 &  25.9  \\
\hline 
\end{tabular}
 \label{tab:cent-event}
\end{table}

\subsection{Track Selection}
The quality of the tracks used in the analysis was assured by employing the standard track selection process used in STAR~\cite{Abelev:2008ab,Adamczyk:2016dzv,Adamczyk:2017iwn}. The track selection criteria are tabulated in Table~\ref{tab:track-cuts}. To avoid intermixing of tracks from a secondary vertex, the track must have a Distance of Closest Approach (DCA) to the primary vertex of less than 3 cm. The number of TPC points used to fit each track (nFitPts) was required to be greater than 25 out of a possible maximum hits (nFitPoss).  The ratio of nFitPts to nFitPoss was required to be greater than 0.52 to exclude split tracks~\cite{thesis}. We also required greater than 15 measured points (ndE/dx) along a track to reliably calculate $\langle dE/dx\rangle$.  And, finally, the rapidity window chosen for this analysis is $|y| < 0.1$ in order to have the largest \pt coverage for particle identification.

\begin{table}
  \centering
	\caption{Track selection criteria for the tracks used in the analysis.}
	\label{tab:track-cuts}\vspace{0.1in}
	\begin{tabular}{lc}
	\hline
	\rule{0pt}{12pt}
	Criterion type & Value \\
	\hline	
	\rule{0pt}{12pt}
 	$|y|$ & $<$ 0.1 \\
	\rule{0pt}{12pt}
	DCA & $<$ 3~cm \\
	\rule{0pt}{12pt}
	nFitPts & $>$ 25 \\
	\rule{0pt}{12pt}
	nFitPts/nFitPoss & $>$ 0.52  \\
	\rule{0pt}{12pt}
	ndE/dx & $>$ 15 \\
	\hline
	\end{tabular}
\end{table}

\subsection{Particle Identification}

\begin{figure*}[!tp]
\centering
\resizebox{0.47\textwidth}{!}{
\includegraphics{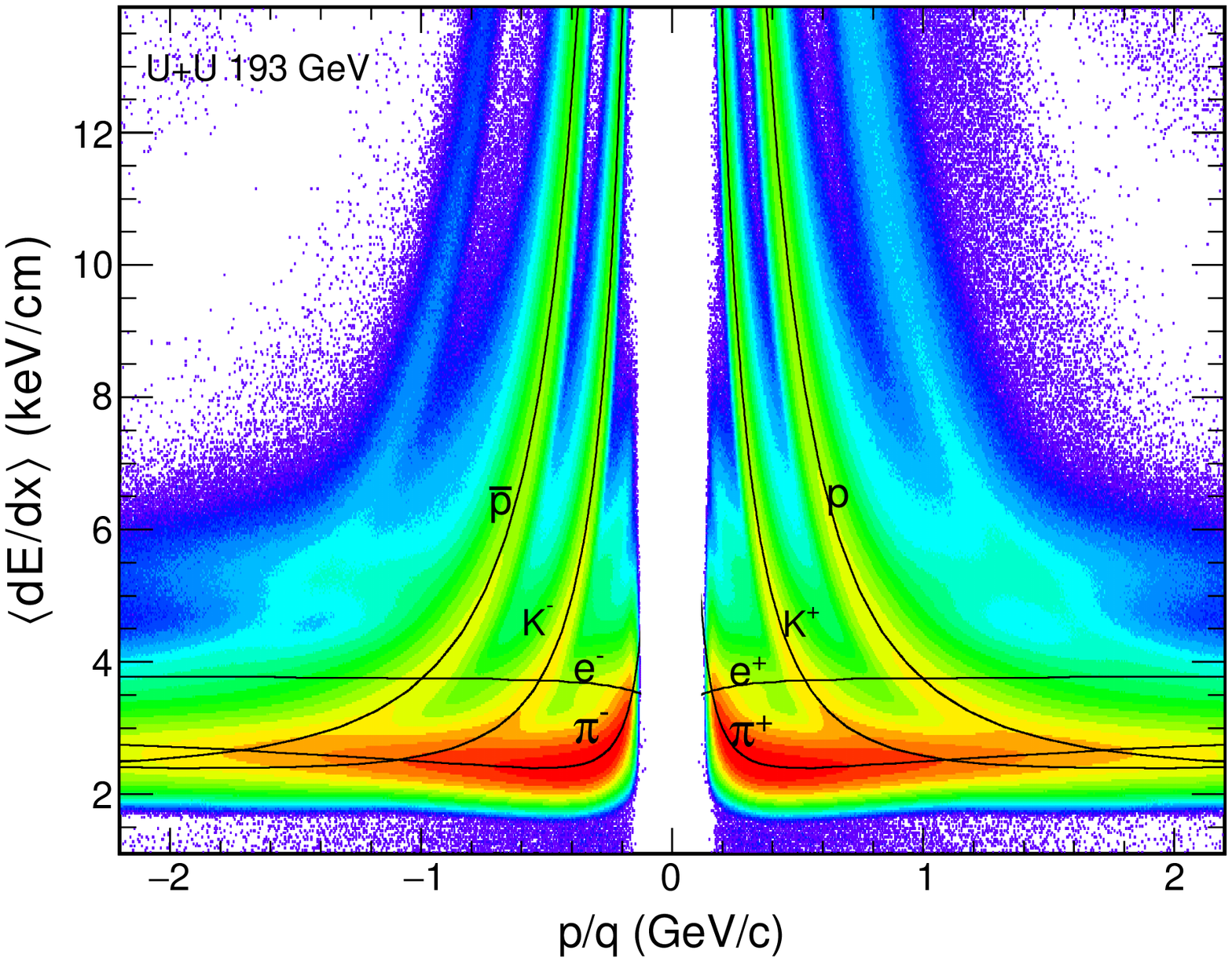}
}
\resizebox{0.45\textwidth}{!}{
\includegraphics{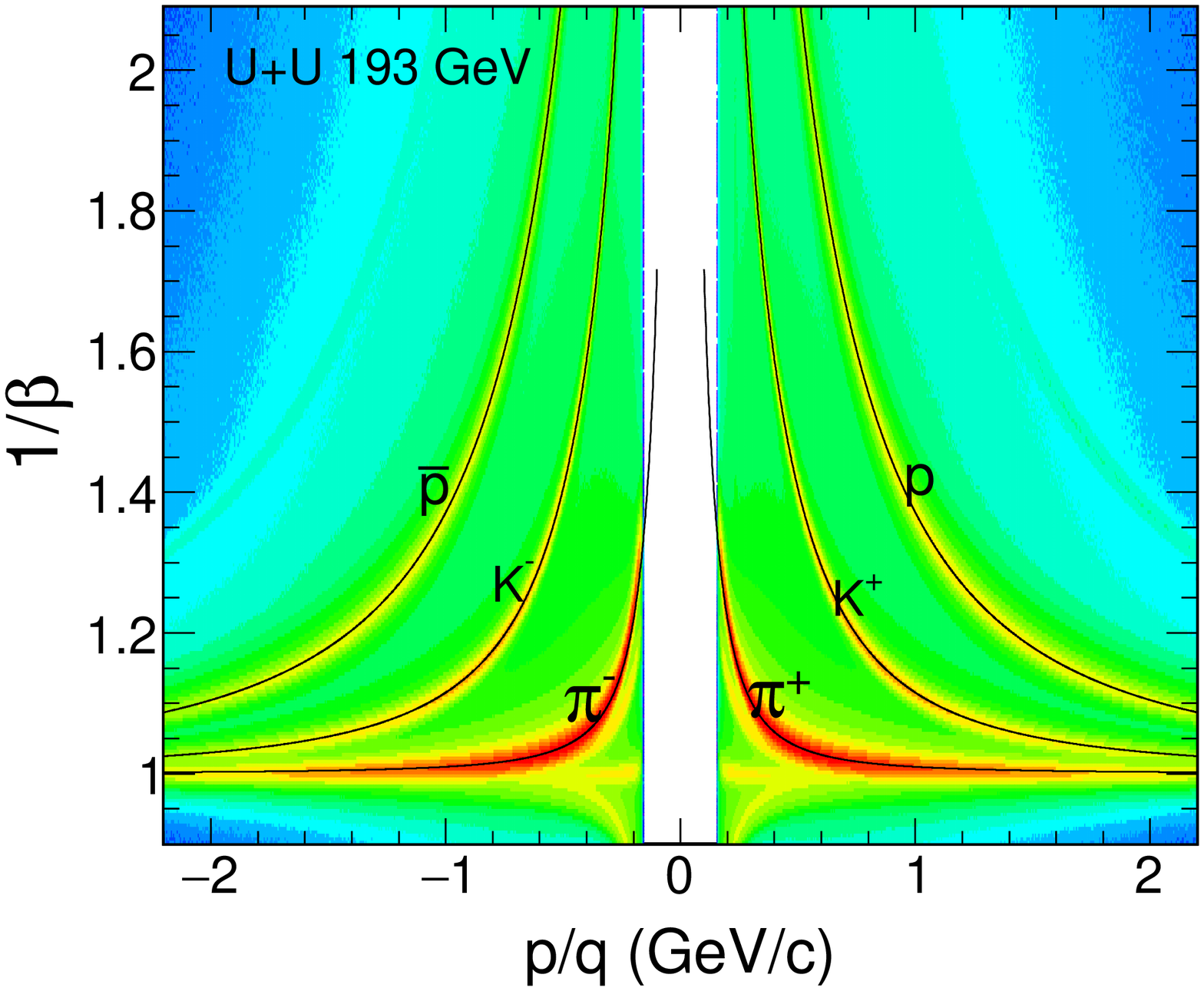}
}
\caption{(Color online) Left panel: The $\langle dE/dx \rangle$ distribution of charged particles as a function of momentum/charge ($p/q$) obtained from TPC in U$+$U collisions at \snn = 193 GeV. The curves represent the expected mean value of $\langle dE/dx \rangle$ for the corresponding particles. Right panel: $1/\beta$ as function of $p/q$. The curves represent the theoretical values of $1/\beta$ for the corresponding particles.}
\label{fig:dEdx-beta-pq}
\end{figure*}

The $\langle dE/dx\rangle$ measured in the TPC  is used for particle identification and is a function of rigidity (momentum-to-charge ratio) as shown in
Fig.~\ref{fig:dEdx-beta-pq} (left panel). The black solid curves represent the theoretical value of $\langle dE/dx\rangle$ predicted by the Bichsel formula~\cite{Bichsel:2006cs}. As visible from the figure, the bands of charged pions, kaons, and protons start merging at higher momentum. Hence, the TPC is extensively used to identify pions, kaons, and protons up to \pt of 0.8, 0.8 and 1.0~GeV/$c$ respectively. 
To extend particle identification to higher \pt we include TOF information for pions and kaons with \pt above 0.4~GeV/$c$ and for protons above 0.5~GeV/$c$. The right side of Fig.~\ref{fig:dEdx-beta-pq} shows the inverse of the particle velocity $1/\beta$ as a function of $p/q$. The solid black lines are the mean values of $1/\beta$ for a particular particle species. The various bands remain clearly separated up to momentum of 2~GeV/$c$. 

\begin{figure*}
\centering
\resizebox{0.92\textwidth}{!}{%
\includegraphics{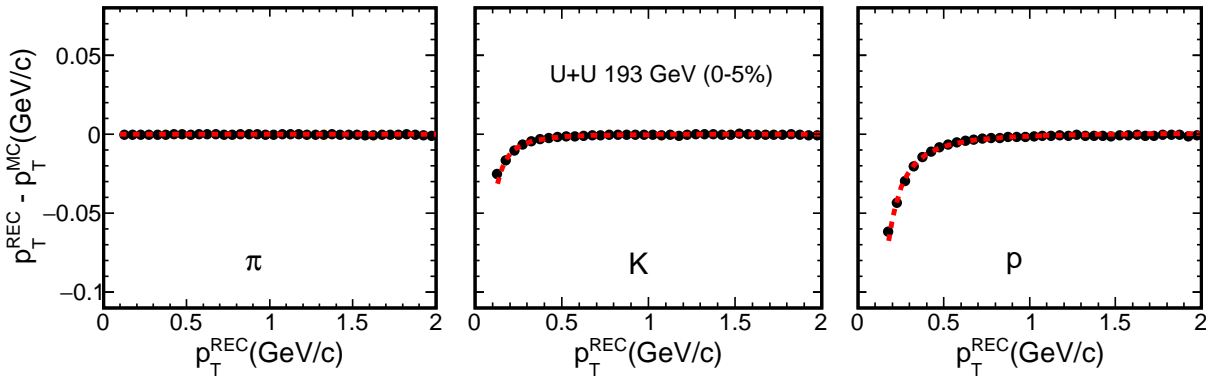}
}
\caption{The difference between $p_{\mathrm T}^{REC}$ and $p_{\mathrm T}^{MC}$ as function of $p_{\mathrm T}^{REC}$ for pions (left), kaons (middle), protons (right) in 0--5\% centrality U+U collisions at \snn = 193~GeV. The errors are statistical only and within symbol size.}
\label{fig:energy-loss}
\end{figure*}

The raw yields of particles identified using $\langle dE/dx\rangle$  from the TPC are obtained using the z-variable~\cite{AguilarBenitez:1991yy} as described in Refs.~\cite{Abelev:2008ab,Adamczyk:2017iwn}. For a particular particle type it is given by the mathematical expression\\
\begin{equation}
z_{X} = \ln \left( \frac{\langle dE/dx \rangle}{\langle dE/dx \rangle_{X}^{B}}\right),
\end{equation}
where $X$ is the particle type chosen [$e^{\pm}$, $\pi^{\pm}$, $K^{\pm}$, $p(\bar{p})$ in the present analysis] and $\langle dE/dx \rangle_{X}^{B}$ is the theoretical value of $\langle dE/dx\rangle$ obtained from Bichsel formula~\cite{Bichsel:2006cs} for the corresponding particle. This gives the $z$-distribution for each particle species within $|y| < 0.1$ for a given \pt and centrality bin.

The $z$ distribution is then fitted by multiple Gaussians, corresponding to each particle species, as discussed in  Ref.~\cite{Adamczyk:2017iwn}. The yields are obtained as the area under the Gaussian function for a particular $p_{T}$ and centrality bin. To extract the raw yields using the TOF, we employ the same technique as described in Refs.~\cite{Adamczyk:2017iwn,Shao:2005iu}. For each detected particle we calculate a parameter ($m^2$) defined as 
\begin{equation}
m^{2} = p^{2}\left( \frac{c^{2}T^{2}}{L^{2}} -1\right),
\end{equation}
where $p$, $T$, $L$, and $c$ respectively represent the particle momentum, time-of-flight, path length of the particle and the velocity of light. The $m^{2}$ distribution in a given \pt and centrality bin is fitted to the predicted $m^2$ distributions as discussed in Refs.~\cite{Adamczyk:2017iwn,Shao:2005iu}.  In this way, the raw yields are extracted in different \pt and centrality bins for each particle species studied here.

\section{CORRECTION FACTORS}
\subsection{Monte-Carlo Embedding Technique}

The Monte-Carlo (MC) embedding technique is briefly described below and can be found in detail in Refs~\cite{Abelev:2008ab,Adamczyk:2017iwn,Abelev:2009bw}. This technique is used to estimate the track reconstruction efficiency and acceptance of the detector.
The MC tracks are generated with flat \pt and $y$ distributions, to ensure equal statistics in each \pt bin. These generated tracks have transverse momentum $p_{\mathrm T}^{MC}$.
The number of MC tracks is about 5\% of  the measured multiplicity in each corresponding $p_{\mathrm T}^{MC}$ bin
and those MC tracks are mixed with the real data tracks at the detector pixel level and allowed to pass through the response of the STAR detector using the GEANT package~\cite{geant}. These events with embedded tracks are reconstructed like real events taking into consideration all the detector effects. 

\subsection{Energy Loss Correction}
In the track reconstruction procedure, the Coulomb scattering and energy loss of a charged particle is corrected by assuming the pion mass for each particle track~\cite{Abelev:2008ab,Adamczyk:2017iwn}. Thus, a momentum correction is needed for higher mass particles such as kaons and protons. This energy loss correction mainly affects the particle yields at low momentum. A correction factor is calculated for kaon and proton tracks from the Monte-Carlo embedding data. The difference between the reconstructed transverse momentum $p_{\mathrm T}^{REC}$ and initial transverse momentum $p_{\mathrm T}^{MC}$ as a function of $p_{\mathrm T}^{REC}$ estimates the energy loss correction track by track. The energy loss correction factor is found to be independent of collision energy and centrality~\cite{Abelev:2008ab,Adamczyk:2017iwn}.
A plot of this energy loss correction as a function of $p_{\mathrm T}^{REC}$ is shown in Fig.~\ref{fig:energy-loss} for this analysis of U+U collisions at \snn = 193 GeV for pions, kaons, and protons. 

\subsection{Tracking Efficiency $\times$ Acceptance}
\begin{figure*}[!tp]
\centering
\resizebox{0.92\textwidth}{!}{%
\includegraphics{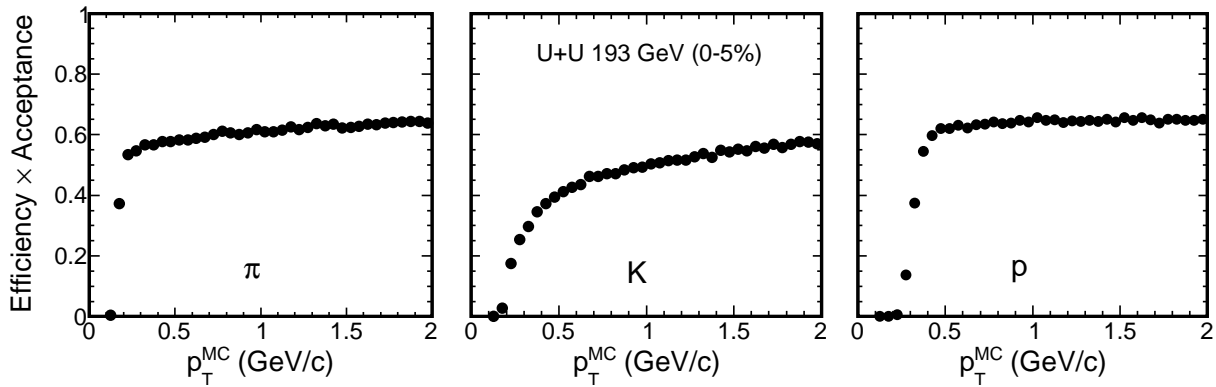}
}
\caption{Tracking efficiency $\times$ acceptance as function of $p_{\mathrm T}^{MC}$ for reconstructed pions (left), kaons (middle), and protons (right) obtained from embedding in U+U collisions at \snn = 193~GeV. The errors are statistical only and within symbol size.}
\label{fig:efficiency}
\end{figure*}

A correction factor taking into account the inefficiency of the detector in reconstructing tracks and its acceptance is applied to the \pt spectra. The MC generated embedding sample provides this correction fraction for each particle species in each centrality. It is given by the ratio of the distribution of reconstructed MC tracks to initial generated MC tracks as a function of $p_{\mathrm T}^{MC}$ in the selected rapidity window. The reconstruction efficiency shows a centrality dependence as it increases from central to peripheral collisions. Figure~\ref{fig:efficiency} represents efficiency $\times$ acceptance as a function of $p_{\mathrm T}^{MC}$ for reconstructed pions, kaons, and protons in 0--5\% centrality U+U collisions at \snn = 193~GeV. The raw \pt spectra of each particle species and centrality are corrected for the efficiency $\times$ acceptance. 

\subsection{TOF Matching efficiency}
The TOF detector surrounds the TPC detector within $|\eta| < 0.9$ and full azimuthal angle. Not all TPC tracks are detected by the TOF detector, hence a track matching correction between TPC and TOF is needed. This correction factor is calculated from data for each particle species. This is obtained from the ratio of the number of tracks detected in the TOF to the number of tracks in the TPC within the acceptance under analysis~\cite{Adamczyk:2017iwn}. This track matching efficiency typically varies from about 60\% (for central collisions) to 70\% (for peripheral collisions) in U+U collisions at \snn = 193~GeV. The \pt spectra in each centrality are divided by this fraction for each \pt bin.

\subsection{Pion Feed-down Correction}
The yield of pions contains contributions from weak decays of $K_{s}^{0}$ and $\Lambda$. Also, muons can be misidentified as pions because of their small mass difference. Hence, a correction factor estimating these weak decay contributions and muon contamination to the pion yield is needed~\cite{Abelev:2008ab,Adamczyk:2017iwn}. This is obtained from MC simulation, where the events generated from Heavy Ion Jet Interaction (HIJING) model~\cite{Wang:1991hta} pass through the STAR detector response using the GEANT package~\cite{geant}. The events are reconstructed in the same manner as that of the real data. In such simulations we have the initial parent particle information. From the final reconstruction, the secondary pions and muons misidentified as pions are calculated. From the final reconstruction we determine the fraction of the pions that are really secondary pions or muons that are misidentified as pions.  Therefore, the pion yields in each \pt bin are reduced by the fraction that are secondary pions and muons identified as pions. The total pion feed-down fraction shows only \pt dependence with no centrality dependence. Its value has a maximum of $15\%$ at \pt = 0.2~GeV/$c$ and vanishes above \pt $>$ 1.2~GeV/$c$. These values are similar to those reported in Refs.\cite{Abelev:2008ab,Adamczyk:2017iwn}.
 
\subsection{Proton Background Correction}
Energetic particles produced in the collision can interact with the beam pipe and detector material producing secondary protons. Because these secondary protons are produced far from the primary vertex, they appear as a long tail in the DCA distribution of protons. However, anti-protons do not have this background and can therefore be used to estimate the proton background fraction by comparing the DCA distributions of proton and anti-proton from the real data~\cite{Abelev:2008ab,Adamczyk:2017iwn,Abelev:2009bw,Adler:2001bp}. In this method, protons and anti-protons are selected within $|n_{\sigma_{p}}| < 2$, where $n_{\sigma_{p}}$ is $z_{X}$ of protons divided by the $\langle dE/dx \rangle$ resolution in the TPC.
The proton background fraction is calculated for each centrality and each \pt bin. The proton background fraction is found to decrease with increasing \pt value. It shows a slight variation with collision centrality (decreases towards central collisions).
The proton background fraction is maximum for low \pt ($\approx$ 9\% for mid-central collisions) and becomes almost negligible for \pt $>$ 1.2 GeV/$c$.
It may be noted that the proton and anti-proton yields reported here are inclusive,  i.e., feed-down contributions from weak decays are not corrected for, similar to those measured in Au+Au collisions at \snn = 200 GeV~\cite{Abelev:2008ab,Adams:2003xp,STAR:2001mal}. 

\section{Systematic Uncertainties}
The systematic uncertainties associated with the results include contributions from the event and track selection criteria, the PID procedure, the methods of estimation of correction factors, and the errors associated with extrapolations with different functional fits to \pt spectra.  The errors associated with all these sources are calculated and are added in quadrature.

The systematic errors associated with event and track selection criteria are estimated by varying one selected criterion while the other criteria are kept at their default values.
The event and track selection criteria with default and varied values are:
30 cm (default) $<$ $V_{z}$ $<$ 40 cm;  2 cm $<$ DCA $<$ 3 cm (default); 20 $<$ nFitPts $<$ 25 (default); and 10 $<$ $ndE/dx$ $<$ 15 (default). In the identification of particle type, the fit range of the Gaussian function for $dE/dx$, and the $n_\sigma$ criteria used in calculation of predicted $m^2$ distribution are varied (from $|n_{\sigma}|<2$ to $|n_{\sigma}|<1$). As another way to obtain the yields, bin counting is used and the resulting difference from the fits is taken as part of the systematic uncertainties.

\begin{table}
  \centering
        \caption{Systematic uncertainties related to the yields of $\pi$, $K$, and $p(\bar{p})$ in U+U collisions at \snn = 193 GeV.}
        \label{tab:tot-sys}\vspace{0.1in}
        \begin{tabular}{lccc}
        \hline  
        ~~
         & $\pi$ & $K$ & $p(\bar{p})$ \\
        \hline
        \rule{0pt}{12pt}
        Cuts and PID & $4\%$ & $5\%$ & $7\%$ \\
        \rule{0pt}{12pt}
        Extrapolation & $8\%$ & $9\%$ & $10\%$ \\          
        \rule{0pt}{12pt}
        Corrections & $5\%$ & $5\%$ & $5\%$ \\
        [3pt]
        \hline 
        \rule{0pt}{12pt}
        Total & $10\%$ & $11\%$ & $13\%$ \\
        \hline
        \end{tabular}
\end{table}

\begin{figure*}[!tp]
\centering
\resizebox{0.99\textwidth}{!}{%
\includegraphics{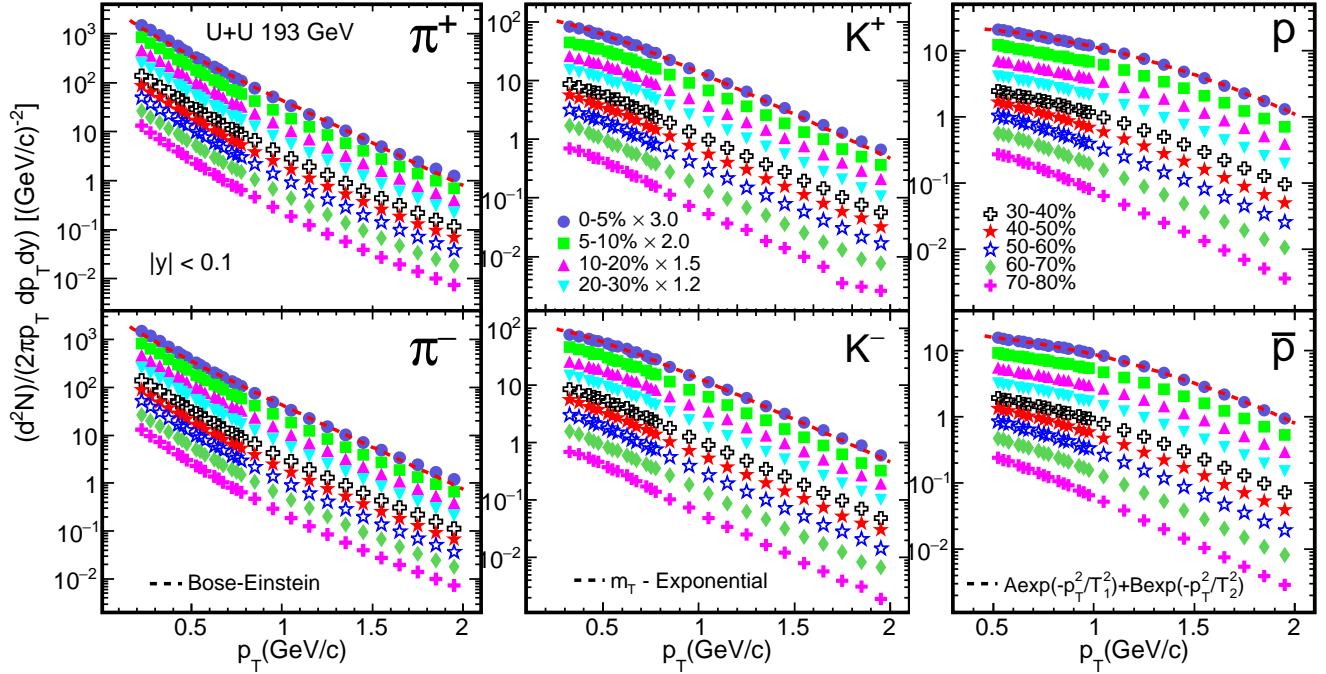}
}
\caption{\pt spectra of $\pi^{\pm}$, $K^{\pm}$, $p(\bar{p})$ measured at midrapidity ($|y| < 0.1$) in U+U collisions at \snn = 193~GeV in STAR. Spectra are plotted for nine different centrality classes and some of them are scaled for clarity. The \pt spectra are fitted, with Bose-Einstein function for pions, $m_{\mathrm T}$-exponential for kaons, and double exponential for (anti) protons, shown for 0--5\% centrality classes in the figure. The uncertainties represent total systematic and statistical uncertainties added in quadrature and are mostly dominated by systematic errors.}
\label{fig:pT-spectra}
\end{figure*}

The procedure for calculation of $dN/dy$ and $\langle p_{\mathrm T}\rangle$ involves the fitting of the spectra by a suitable function to extrapolate the spectra to the unmeasured \pt region. The main source of uncertainty on $dN/dy$ and $\langle p_{\mathrm T}\rangle$ comes from this extrapolation. In order to estimate the uncertainty associated with extrapolation, different fit functions are used. The default fit functions used for pions, kaons, and protons are respectively Bose-Einstein, $m_{\mathrm T}$~exponential and double exponential respectively~\cite{Adamczyk:2017iwn}. For the systematic uncertainty estimation the used functions are \pt exponential, Boltzmann, and $m_{\mathrm T}$~exponential for pions, kaons, and protons, respectively~\cite{Adamczyk:2017iwn}. The systematic uncertainties on the values of $\langle p_{\mathrm T} \rangle$ for pions, kaons, and protons are estimated to be 9\%, 10\% and 10\%, respectively.

In addition, a 5\% estimated systematic uncertainty associated with the calculation of the detector efficiency and acceptance was added in quadrature to the statistical errors on the \pt spectra~\cite{Abelev:2008ab}. The systematic uncertainties associated with the pion feed-down and proton background correction were also estimated.  While the contribution from the former is negligible, the latter adds a 6--7\% systematic uncertainty to the low \pt part of the proton \pt spectra.  The different sources of systematic uncertainties on the particle yields for all centralities are tabulated in Table~\ref{tab:tot-sys}. 

The systematic uncertainty  on the particle ratios are obtained using the systematic uncertainty on the particle yields.
The correlated uncertainties, i.e., from efficiency, will cancel out completely in the particle ratios. The systematic uncertainties associated with the extrapolated fits also cancels out for particle and anti-particle ratios.

The uncertainties associated with the kinetic freeze-out parameters have two contributions which are added in quadrature. One of them is the uncertainty that is obtained during the simultaneous fitting of pion, kaon, and proton spectra along with their point-to-point systematic uncertainties. The second is obtained by varying the fit range of the measured \pt spectra.

\section{RESULTS AND DISCUSSIONS}
\subsection{Transverse Momentum Spectra}
The transverse momentum spectra in U+U collisions at \snn = 193~GeV for $\pi^{+}$, $\pi^{-}$, $K^+$, $K^-$, $p$ and $\bar{p}$  within $|y|< 0.1$ are shown in Fig.~\ref{fig:pT-spectra}. The spectra are shown in nine different centrality classes 0--5\%, 5--10\%, 10--20\%, 20--30\%, 30--40\%, 40--50\%, 50--60\%, 60--70\% and 70--80\%. The data points shown for 0.4 $<$ \pt $<$  2.0 GeV/$c$ for pions and kaons and for 0.5 $<$ \pt $<$  2.0 GeV/$c$ for protons are obtained using both the TPC and the TOF. In addition, the data points measured using only the TPC are shown for  0.2 $<$ \pt $<$ 0.8 GeV/$c$, 0.3 $<$ \pt $<$ 0.8 GeV/$c$ and 0.5 $<$ \pt $<$ 1.0~GeV/$c$ regions for pions, kaons, and protons respectively. The 0.4 $<$ \pt $<$ 0.8 GeV/$c$, 0.4 $<$ \pt $<$ 0.8 GeV/$c$ and 0.5 $<$ \pt $<$ 1.0 GeV/$c$ for pions, kaons, and protons respectively are the overlap region containing data points measured using only the TPC and also the data points using the TPC along with the TOF.
For each particle in Fig.~\ref{fig:pT-spectra} the spectral shape in this overlapping \pt range is seen to smoothly join the data obtained using only the TPC (low \pta) and the data obtained using the TOF (high \pta).
The curves represent the fit to the spectra, shown only for 0--5\% central collisions. The respective functions for pion, kaons and protons are Bose-Einstein, $m_{\mathrm T}$~exponential and double exponential~\cite{Adamczyk:2017iwn}. The particle yields ($dN/dy$) and average transverse momenta ($\langle p_{\mathrm T}\rangle$) are obtained from the measured points in the \pt spectra. For the unmeasured \pt regions, the contributions to the yield are extracted by extrapolating the fit functions. The behavior of the \pt spectra changes with centrality. As can be observed from the Fig.~\ref{fig:pT-spectra}, the slopes of the spectra show a gradual flattening as one goes from peripheral to central collisions. This is an indication of stronger radial flow effects for particles with increasing centrality. A mass dependence of the inverse slope of the \pt spectra can also be inferred as it increases with increasing mass of the particle.

\begin{figure*}[!tp]
\centering
\resizebox{1.0\textwidth}{!}{%
  \includegraphics{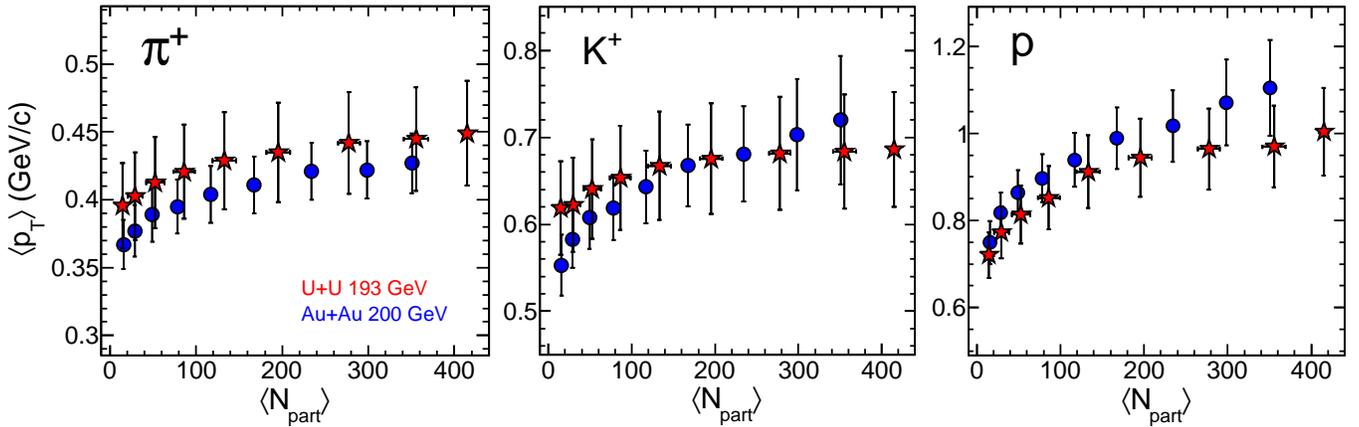}
}
\caption{$\langle p_{\mathrm T}\rangle$ of $\pi^{+}$, $K^{+}$ and $p$ as a function of $\langle N_{\text{part}}\rangle$ at midrapidity ($|y| < 0.1$) for U+U collisions at \snn = 193~GeV. The results are compared with Au+Au collisions at \snn = 200~GeV~\cite{Abelev:2008ab}.
The uncertainties represent the total systematic and statistical uncertainties added in quadrature, and are dominated by systematic uncertainties. }
\label{fig:mean-pT}
\end{figure*}

\subsection{Average Transverse Momenta}
The slopes of the spectra can be quantified by $\langle p_{\mathrm T}\rangle$.
The $\langle p_{\mathrm T}\rangle$ for $\pi^{+}$, $K^{+}$ and $p$ is plotted as a function of $\langle N_{\text{part}}\rangle$ in Fig.~\ref{fig:mean-pT}  for U+U collisions at \snn = 193~GeV.
For comparison,  the published results obtained in Au+Au collisions at \snn = 200~GeV are also shown ~\cite{Abelev:2008ab}.
Not shown explicitly are the $\langle p_{\mathrm T}\rangle$ for $\pi^{-}$, $K^{-}$ and $\bar{p}$ which show identical behavior. The $\langle p_{\mathrm T}\rangle$ increases gradually from peripheral to central collisions as well as from pions to kaons and kaons to protons in U+U collisions at \snn = 193~GeV. The hadron mass dependence of $\langle p_{\mathrm T}\rangle$ is indicates radial flow effect, which increases from peripheral to central collisions. The $\langle p_{\mathrm T}\rangle$ are similar for collisions between deformed uranium nuclei and spherical gold nuclei for comparable values of $\langle N_{\text{part}} \rangle$. Moreover, using U+U collisions the $\langle p_{\mathrm T}\rangle$ measurements have been made for the highest $\langle N_{\text{part}} \rangle$ values observed at RHIC. The $\langle p_{\mathrm T}\rangle$ values are listed in Table~\ref{tab:meanpt} for U+U collisions at \snn = 193~GeV.

\begin{table*}{}
\caption{$\langle p_{\mathrm T} \rangle$ values in GeV/$c$ for $\pi^{+}$, $\pi^{-}$,$K^{+}$, $K^{-}$ $p$, and $\bar{p}$ from U+U collisions at \snn = 193~GeV within rapidity $|y| < 0.1$. The quoted errors are the total systematic and statistical errors added in quadrature, which are dominated by the systematic errors.}
\label{tab:meanpt}
\begin{tabular}{ccccccc} 
\hline 
  Centrality (\%) & $\pi^{+}$  & $\pi^{-}$ & $K^{+}$ & $K^{-}$ & $p$ & $\bar{p}$ \\ 
\hline 
	$0-5$ &0.449~$\pm$~0.039~~&~~0.446~$\pm$~0.038 ~&~ 0.686~$\pm$~0.066 ~&~ 0.685~$\pm$~0.065 ~&~ 1.004~$\pm$~0.101 ~&~ 1.003~$\pm$~0.101 \\
	$5-10$ & 0.445~$\pm$~0.038 ~&~ 0.443~$\pm$~0.038 ~&~ 0.684~$\pm$~0.065 ~&~ 0.682~$\pm$~0.065 ~&~ 0.969~$\pm$~0.094 ~&~ 0.961~$\pm$~0.092 \\
	$10-20$ & 0.442~$\pm$~0.038 ~&~ 0.439~$\pm$~0.037 ~&~ 0.682~$\pm$~0.065 ~&~ 0.680~$\pm$~0.065 ~&~ 0.964~$\pm$~0.093 ~&~ 0.963~$\pm$~0.093 \\
	$20-30$ & 0.435~$\pm$~0.037 ~&~ 0.432~$\pm$~0.036 ~&~ 0.676~$\pm$~0.064 ~&~ 0.677~$\pm$~0.064 ~&~ 0.944~$\pm$~0.089 ~&~ 0.943~$\pm$~0.089 \\
	$30-40$ & 0.429~$\pm$~0.036 ~&~ 0.427~$\pm$~0.036 ~&~ 0.667~$\pm$~0.062 ~&~ 0.666~$\pm$~0.061 ~&~ 0.912~$\pm$~0.083 ~&~ 0.911~$\pm$~0.083 \\
	$40-50$ & 0.421~$\pm$~0.035 ~&~ 0.414~$\pm$~0.034  ~&~ 0.654~$\pm$~0.060 ~&~ 0.656~$\pm$~0.060 ~&~ 0.852~$\pm$~0.073 ~&~ 0.852~$\pm$~0.073 \\
	$50-60$ & 0.413~$\pm$~0.034 ~&~ 0.404~$\pm$~0.033 ~&~ 0.640~$\pm$~0.057 ~&~ 0.643~$\pm$~0.058 ~&~ 0.814~$\pm$~0.066 ~&~ 0.813~$\pm$~0.066 \\
	$60-70$ & 0.403~$\pm$~0.032 ~&~ 0.398~$\pm$~0.032  ~&~ 0.622~$\pm$~0.054 ~&~ 0.623~$\pm$~0.063 ~&~ 0.773~$\pm$~0.060 ~&~ 0.760~$\pm$~0.058 \\
	$70-80$ & 0.396~$\pm$~0.031 ~&~ 0.390~$\pm$~0.030 ~&~ 0.619~$\pm$~0.054 ~&~ 0.617~$\pm$~0.053 ~&~ 0.721~$\pm$~0.052 ~&~ 0.702~$\pm$~0.049 \\
	\hline 
\end{tabular} 
 \end{table*}

\begin{figure*}[!tp]
\centering
\resizebox{0.65\textwidth}{!}{%
\includegraphics{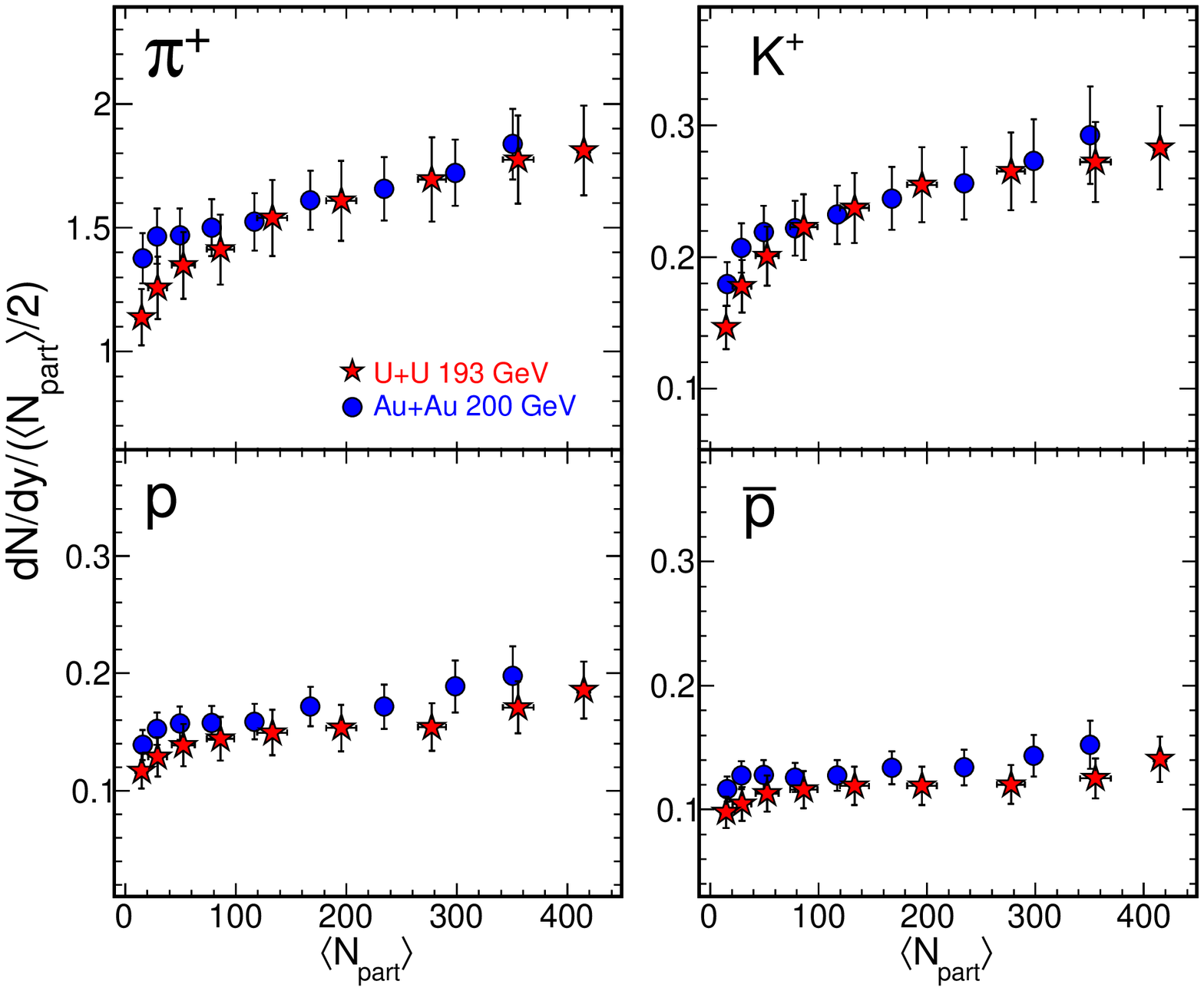}
}
\caption{$dN/dy$ of $\pi^{+}$, $K^{+}$, $p$ and $\bar{p}$ scaled by $\langle N_{\text{part}}\rangle/2$ as a function of $\langle N_{\text{part}}\rangle$ at midrapidity ($|y| < 0.1$) for U+U collisions at \snn = 193~GeV. The results are compared with Au+Au collisions at \snn = 200~GeV~\cite{Abelev:2008ab}.
The uncertainties represent total systematic and statistical uncertainties added in quadrature, dominated by systematic uncertainties.}
\label{fig:dNdy}
\end{figure*}
\subsection{Particle Yields}
The total yield of a particle in a given centrality is obtained by integrating the spectra over the full \pt range. The yields from the  unmeasured region of the spectra are extracted using the functional extrapolation~\cite{Adamczyk:2017iwn}. 
It is represented by the observable $dN/dy$ within rapidity $|y| < 0.1$. The normalised $dN/dy$ with $\langle N_{\text{part}}\rangle/2$ of $\pi^{+}$, $K^{+}$, $p$ and $\bar{p}$ measured in U+U collisions at \snn = 193~GeV as a function of $\langle N_{\text{part}}\rangle$ is shown in Fig.~\ref{fig:dNdy}. The results are compared with the published results of Au+Au collisions at \snn = 200~GeV~\cite{Abelev:2008ab}.
The values of $dN/dy$ are listed in Table~\ref{tab:dndy} for U+U collisions at \snn = 193~GeV.

The current data extends the measurements to a higher value of $N_{\text{part}}$, with $N_{\text{part}}$ extending above 400. Figure~\ref{fig:dNdy} shows that the values of $dN/dy$ per participating nucleon pair for $\pi^{+}$, $K^{+}$ and $p$ increase with average number of participant nucleons in U+U collisions at \snn = 193~GeV. Not shown are the plots for $\pi^{-}$ and $K^{-}$ that show identical behavior to $\pi^{+}$ and $K^{+}$. 
This signifies particle production having contributions from both soft processes and hard processes involving nucleon-nucleon binary collisions. However the $N_{\text{part}}$ dependence is smaller for
anti-protons, possibly due to annihilation effects. The values of $dN/dy$ are consistent between U+U collisions at \snn = 193~GeV and Au+Au collisions at \snn = 200~GeV within a similar range of $\langle N_{\text{part}}\rangle$ and within the quoted systematic uncertainties. 
The comparison between U+U and Au+Au collision systems was also reported in Ref.~\cite{PHENIX:2020lix} on the production of $\pi^0$ and $\eta$, where it was found that the yields of these particles show similar suppression in U+U and Au+Au collisions relative to the binary scaled $p+p$ collisions.

\begin{table*}{}
\caption{$dN/dy$ values for $\pi^{+}$, $\pi^{-}$,$K^{+}$, $K^{-}$ $p$, and $\bar{p}$ from U+U collisions at \snn = 193~GeV within rapidity $|y| < 0.1$. The quoted errors are the total systematic and statistical errors added in quadrature, which are dominated by the systematic errors.}
\label{tab:dndy}
\begin{tabular}{ccccccc} 
\hline 
  Centrality (\%) & $\pi^{+}$ & $\pi^{-}$ & $K^{+}$ & $K^{-}$ & $p$ & $\bar{p}$ \\ 
\hline 
	$0-5$ & 375.9~$\pm$~37.6 ~&~ 377.9~$\pm$~37.8 ~&~ 58.3~$\pm$~6.5 ~&~ 54.7~$\pm$~6.1 ~&~ 38.5~$\pm$~5.0 ~&~ 29.2~$\pm$~3.8 \\
	$5-10$ & 315.4~$\pm$~31.6 ~&~ 316.0~$\pm$~31.6 ~&~ 48.0~$\pm$~5.4 ~&~ 46.0~$\pm$~5.2 ~&~ 30.4~$\pm$~4.0 ~&~ 22.3~$\pm$~2.9\\
	$10-20$ & 235.1~$\pm$~23.5 ~&~ 236.4~$\pm$~23.7 ~&~ 36.2~$\pm$~4.1 ~&~ 34.4~$\pm$~3.8 ~&~ 21.4~$\pm$~2.8 ~&~ 16.7~$\pm$~2.2\\
	$20-30$ & 157.1~$\pm$~15.7 ~&~ 158.9~$\pm$~15.9 ~&~ 24.9~$\pm$~2.8 ~&~ 23.5~$\pm$~2.6 ~&~ 15.0~$\pm$~2.0 ~&~ 11.7~$\pm$~1.5\\
	$30-40$ & 102.4~$\pm$~10.2 ~&~ 103.3~$\pm$~10.3 ~&~ 15.8~$\pm$~1.8 ~&~ 15.3~$\pm$~1.7 ~&~ 9.94~$\pm$~1.29 ~&~ 7.92~$\pm$~1.03\\
	$40-50$ & 60.9~$\pm$~6.1 ~&~ 61.9~$\pm$~6.2  ~&~ 9.59~$\pm$~1.07 ~&~ 9.13~$\pm$~1.02 ~&~ 6.21~$\pm$~0.81 ~&~ 5.00~$\pm$~0.65\\
	$50-60$ & 35.5~$\pm$~3.6 ~&~ 36.6~$\pm$~3.6 ~&~ 5.28~$\pm$~0.59 ~&~ 4.98~$\pm$~0.56 ~&~ 3.65~$\pm$~0.47 ~&~ 2.97~$\pm$~0.39\\
	$60-70$ & 18.5~$\pm$~1.9 ~&~ 18.7~$\pm$~1.9 ~&~ 2.59~$\pm$~0.29 ~&~ 2.37~$\pm$~0.27 ~&~ 1.89~$\pm$~0.25 ~&~ 1.54~$\pm$~0.20\\
	$70-80$ & 8.35~$\pm$~0.84 ~&~ 8.51~$\pm$~0.86 ~&~ 1.07~$\pm$~0.12 ~&~ 1.01~$\pm$~0.11  ~&~ 0.856~$\pm$~0.111 ~&~ 0.717~$\pm$~0.093 \\
	\hline 
\end{tabular} 
\end{table*}

\begin{figure*}[!tp]
\centering 
\resizebox{1.0\textwidth}{!}{%
 \includegraphics{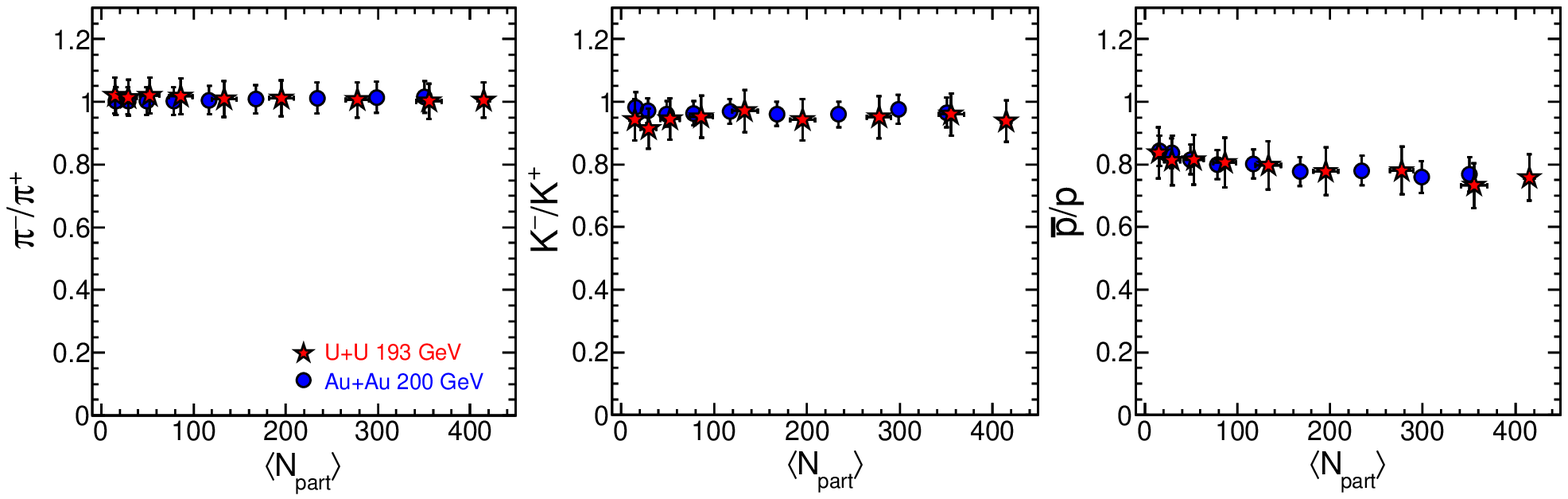}
}
\caption{$\pi^{-}/ \pi^{+}$, $K^{-}/K^{+}$ and $\bar{p}/p$ ratios at midrapidity ($|y| < 0.1$) as a function of $\langle N_{\text{part}}\rangle$ in U+U collisions at \snn = 193~GeV. The results are compared with Au+Au collisions at \snn = 200~GeV~\cite{Abelev:2008ab}.
The uncertainties represent total systematic and statistical uncertainties added in quadrature, dominated by systematic uncertainties.}
\label{fig:like-ratios}
\end{figure*}
\subsection{Particle Ratios}
\begin{figure*}[!tp]
\centering
\resizebox{0.65\textwidth}{!}{%
    \includegraphics{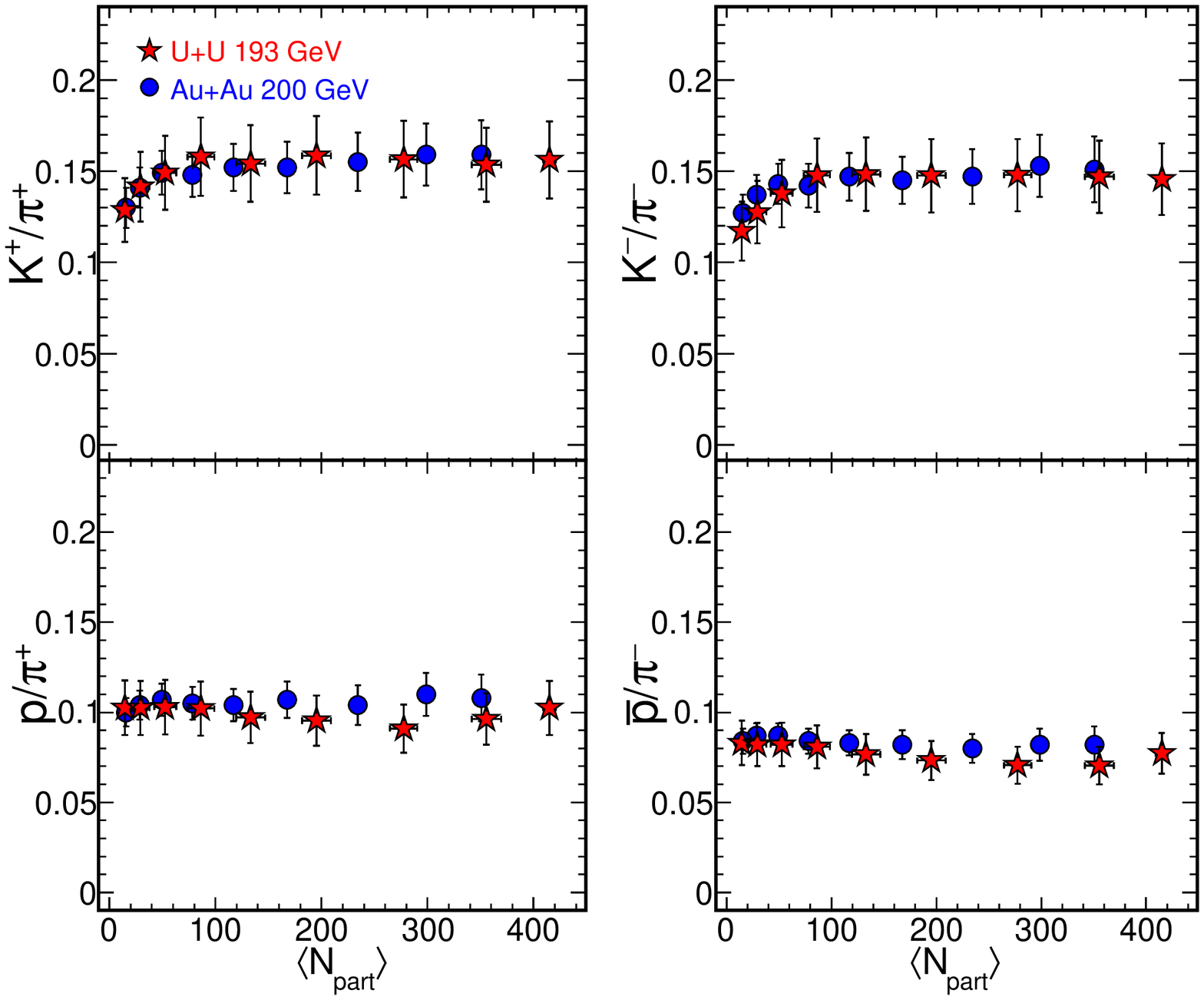}
}
\caption{$K^{+}/ \pi^{+}$, $K^{-}/ \pi^{-}$, $p/ \pi^{+}$ and $\bar{p}/\pi^{-}$
particle yield ratios at midrapidity  ($|y| < 0.1$) as a function of $\langle N_{\text{part}}\rangle$ in U+U collisions at \snn = 193~GeV. The results are compared with Au+Au collisions at \snn = 200~GeV~\cite{Abelev:2008ab}.
The uncertainties represent the total systematic and statistical uncertainties added in quadrature, dominated by systematic uncertainties.}
\label{fig:unlike-ratios}
\end{figure*} 
 
Particle yield ratios are measured in U+U collisions at \snn = 193~GeV and compared with the published results from Au+Au collisions at \snn = 200~GeV~\cite{Abelev:2008ab}. 
Ratios of anti-particle to particle yields $\pi^{-}/ \pi^{+}$, $K^{-}/K^{+}$ and $\bar{p}/p$ as a function of $\langle N_{\text{part}}\rangle$ in U+U collisions at \snn = 193~GeV are shown in Fig.~\ref{fig:like-ratios}. As a function of $\langle N_{\rm{part}}\rangle$, the yield ratios shown are all approximately constant and near unity. For the $\pi^{-}/\pi^{+}$ yield ratios, this is an indication that the particle production mechanism for $\pi^{+}$ and $\pi^{-}$ does not change significantly from central to peripheral collisions. For the $K^{-}/K^{+}$ yield ratios, this suggests that $K^{+}$ and $K^{-}$ are dominantly produced through a pair production process.
The $\bar{p}/p$ yield ratios show a slight increase as the collisions change from central to peripheral, which may be an indication that the baryon stopping power is decreasing toward more peripheral collisions. Within systematic uncertainties, these yield ratios agree with the published results in Au+Au collisions at \snn = 200 ~GeV~\cite{Abelev:2008ab}.

Unlike particle yield ratios for $K^{+}/ \pi^{+}$, $K^{-}/ \pi^{-}$, $p/ \pi^{+}$ and $\bar{p}/ \pi^{-}$ as a function of $\langle N_{\text{part}}\rangle$ in U+U collisions at \snn = 193~GeV are presented in Fig.~\ref{fig:unlike-ratios}. Earlier published results from STAR in Au+Au collisions at \snn = 200~GeV~\cite{Abelev:2008ab}
are also shown for comparison. The $K^{+} / \pi^{+}$ and $K^{-} / \pi^{-}$ yield ratios gradually increase from peripheral to mid-central collisions and saturate from mid-central to central collisions. It could be due to strangeness equilibrium described in various thermodynamical models~\cite{Kaneta:2004zr,Cleymans:2004pp} as well as due to baryon stopping at midrapidity~\cite{Wang:1999rf,Wang:1999xi,BraunMunzinger:2001as}. 
The $p/\pi^{+}$ and $\bar{p}/\pi^{-}$ yield ratios are essentially independent of collision centrality in U+U collisions at \snn = 193~GeV. 
The behavior of all these unlike particle yield ratios in U+U collisions at \snn = 193~GeV is similar, within the quoted uncertainties, to those from the Au+Au collisions at \snn = 200~GeV~\cite{Abelev:2008ab}.

\subsection{Kinetic Freeze-out}
The evolution of the system with time in high-energy heavy-ion collisions takes the system through different stages.
Typically these stages are the QGP phase, the phase transition/crossover, the hadron gas phase, the chemical freeze-out, and the kinetic freeze-out.  This last stage fixes the particles’ momenta and they stream freely toward the detector.  This stage is characterized by the kinetic freeze-out temperature $T_{k}$ and radial flow velocity $\beta$.
\begin{figure}
\centering
\includegraphics[width=0.4 \textwidth] {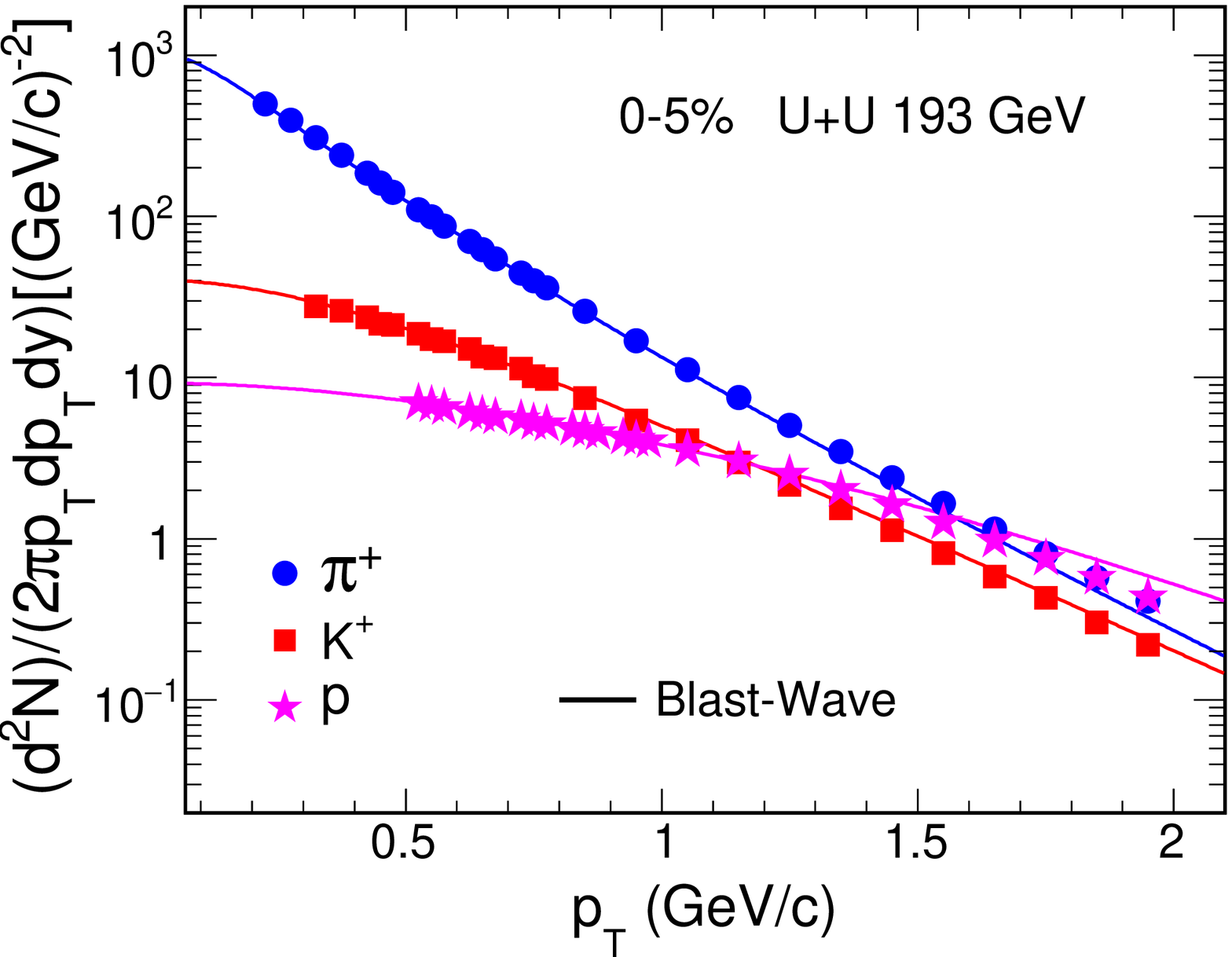}
\caption{Simultaneous Blast-wave fit to the $\pi^{\pm}$, $K^{\pm}$, $p(\bar{p})$ \pt spectra in U+U collisions at \snn = 193 GeV in 0-5\% centrality class within $|y| < 0.1$. For clarity only distributions for positively charged particles are shown. The uncertainties represent the total systematic and statistical uncertainties added in quadrature.}
\label{fig:bw-fit}
\end{figure}

\begin{figure*}[!tp]
\centering
\includegraphics[width=1.0\textwidth]{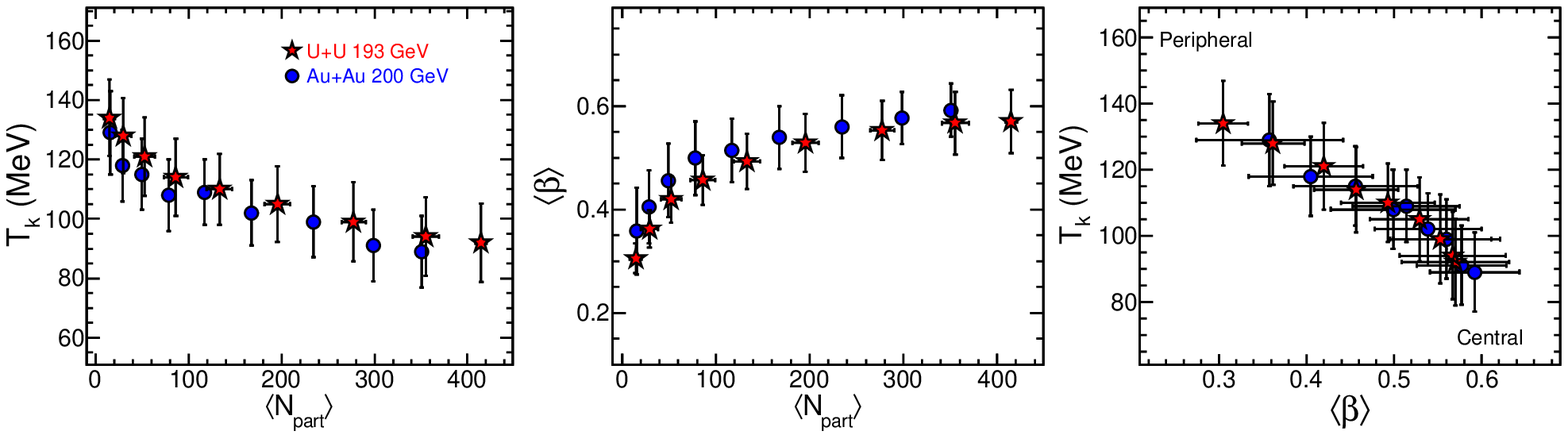}
\caption{Left panel: $T_{k}$ as a function of $\left< N_{\rm part} \right>$. Middle panel: $\langle \beta \rangle$ as a function of $\left< N_{\rm part} \right>$. Right panel: variation of $T_{k}$ with $\langle \beta \rangle$. The midrapidity ($|y| < 0.1$) results from U+U collisions at \snn =  193~GeV  are compared with Au+Au collisions at \snn = 200~GeV~\cite{Abelev:2008ab}.
The uncertainties represent the total systematic and statistical uncertainties added in quadrature, dominated by systematic uncertainties.}
\label{fig:kfo-para}
\end{figure*}

The kinetic freeze-out parameters are usually calculated from the hydrodynamics based Blast-Wave model~\cite{Schnedermann:1993ws}. This model assumes at each point in time a locally thermalised system with temperature $T_{k}$ and moving with a common radial velocity $\beta$. The transverse distribution of particles for such a radially boosted system can be written as,
\begin{eqnarray}
\frac{dN}{p_{\mathrm T}dp_{\mathrm T}} \propto \int_{0}^{R} r dr \,m_{\mathrm T} I_{0} \left( \frac{p_{\mathrm T}\sinh\rho(r)}{T_{k}} \right) \nonumber \\
 \times K_{1} \left( \frac{m_{\mathrm T}\cosh\rho(r)}{T_{k}} \right), \label{eq:bw}
\end{eqnarray}
where transverse mass $m_{\mathrm T} = \sqrt{p_{\mathrm T}^{2} + m^{2}}$, $m$ being the mass of the hadron; $\rho(r) = \tanh^{-1}\beta$; $I_{0}$ and $K_{1}$ are the modified Bessel functions. We use a radial flow velocity profile of the form \\
\begin{equation}
\beta = \beta_{s} (r/R)^n,
\end{equation}
where $\beta_{s}$ is the surface velocity and $r/R$ is the relative radial position in the thermal source with $n$ being the exponent of the flow velocity profile. The average radial flow velocity $\langle \beta \rangle $ can be calculated as $\langle  \beta \rangle  = \frac{2}{2+n} \beta_{s}$. 

Using the data shown in Fig.~\ref{fig:pT-spectra}, the kinetic freeze-out parameters are extracted separately for the data from each centrality class by applying a simultaneous Blast-wave fit~\cite{Adamczyk:2017iwn,Abelev:2009bw} to the $\pi^{\pm}$, $K^{\pm}$ and $p(\bar{p})$ spectra for a given centrality class.  The results for one such fit to the 0-5\% centrality class are shown in Fig.~\ref{fig:bw-fit}, illustrated as overlaid on the $\pi^{+}$, $K^{+}$ and $p$  spectra.  Several general observations can be made.   The Blast-wave model is very sensitive to the \pt~range of the spectrum used in the fit~\cite{Abelev:2013vea}.  Specifically, the low \pt~part of the pion spectra are mostly affected by resonance decays. Thus, the pion spectra are fitted for \pt $>$ 0.5~GeV/$c$.  In addition, the Blast-wave model does not fit the high \pt~part of the spectra~\cite{Wilk:1999dr}.  

The kinetic freeze-out parameters obtained from the Blast-wave fits to the U+U collisions at \snn = 193~GeV are compared with those reported for the Au+Au collisions at \snn = 200 GeV for a similar \pt~range ~\cite{Adamczyk:2017iwn, Abelev:2008ab} and are shown in Fig.~\ref{fig:kfo-para}. 
As mentioned earlier, the pion spectra are fitted  for \pt greater than 0.5 GeV/$c$, while kaon and proton spectra are fitted from the lowest available \pt. Typically high \pt values are limited below 1.2 GeV/$c$.  
From the left panel showing $T_{k}$  vs. $\langle N_{\text{part}}\rangle$ it can be seen that $T_{k}$ decreases with $\langle N_{\text{part}}\rangle$,
while the middle panel shows $\langle \beta\rangle$ increasing with $\langle N_{\text{part}}\rangle$.
This suggests higher rate of expansion as the collision centrality increases. For peripheral collisions, this suggest a shorter-lived fireball~\cite{Heinz:2004qz}.
The right panel shows the correlation between $T_{k}$ and $\langle \beta\rangle$ such that an increase in $T_{k}$ corresponds to a decrease in $\langle \beta\rangle$ and vice-versa.  In all three panels it is seen that the values obtained from U+U collisions at \snn = 193~GeV are consistent within errors with those reported for the Au+Au collisions at \snn = 200 GeV~\cite{Abelev:2008ab}.  The kinetic freeze-out parameters in U+U collisions at \snn = 193~GeV are listed in Table~\ref{tab:kfo-para}.

\begin{table}[!h]
  \centering 
  \caption{Kinetic freeze-out parameters in U+U collisions at \snn = 193 GeV. The quoted errors are statistical and systematic errors added in quadrature.}
  \label{tab:kfo-para}\vspace{0.1in}
  \begin{tabular}{ccccc}
    \hline	
    Centrality & $T_{k}$ (MeV) & $\left\langle  \beta \right\rangle$  & n & $\chi^{2}/ndf$ \\ 
    \hline 
    $0-5\%$ & 92~$\pm$~12 ~&~ 0.570~$\pm$~0.041 ~&~ 0.970$~\pm$~0.074 ~&~ 0.097 \\
    $5-10\%$ & 94~$\pm$~12 ~&~ 0.567~$\pm$~0.040 ~&~ 0.947~$\pm$~0.152 ~&~ 0.043 \\
    $10-20\%$ & 99~$\pm$~12 ~&~ 0.553~$\pm$~0.037 ~&~ 0.987~$\pm$~0.153 ~&~ 0.042 \\
	  $20-30\%$ & 105~$\pm$~11 ~&~ 0.529~$\pm$~0.037 ~&~ 1.035~$\pm$~0.183 ~&~ 0.049 \\
	  $30-40\%$ & 110~$\pm$~10& 0.493~$\pm$~0.036 ~&~ 1.180~$\pm$~0.210 ~&~ 0.054 \\
	  $40-50\%$ & 114~$\pm$~11 ~&~ 0.457~$\pm$~0.031 ~&~ 1.449~$\pm$~0.229 ~&~ 0.052 \\
	  $50-60\%$ & 121~$\pm$~11 ~&~ 0.420~$\pm$~0.030 ~&~ 1.564~$\pm$~0.272 ~&~ 0.061 \\
	  $60-70\%$ & 128~$\pm$~10 ~&~ 0.362~$\pm$~0.021 ~&~ 2.292~$\pm$~0.269 ~&~ 0.062 \\
	  $70-80\%$ & 134~$\pm$~10 ~&~ 0.305~$\pm$~0.015 ~&~ 3.158~$\pm$~0.554 ~&~ 0.102 \\
	 \hline 
	\end{tabular}
\end{table}

\subsection{AMPT Model Comparison}
In this section, a detailed comparison of the STAR results for U+U collisions at \snn = 193~GeV is done with the AMPT model calculations (version 2.25t7d and 2.26t9)~\cite{Lin:2004en}. The comparison of the AMPT model with results from Au+Au collisions at \snn = 200~GeV has been done in Ref.~\cite{Nandi:2019ztz}.
Such a comparison allows us to constrain the relevant partonic cross sections for the specific reactions. 

In the AMPT model, the initial particle distribution is generated by the HIJING model~\cite{Wang:1991hta}. For the results presented here, the string melting version of the AMPT model is used which incorporated partonic and hadronic interactions. To incorporate the deformation of uranium nuclei, the AMPT model has been modified, as described in Ref.~\cite{Haque:2011aa} and briefly discussed below. The nucleon density distribution is parametrized as a deformed Woods-Saxon profile~\cite{Hagino:2006fj}
\begin{eqnarray}
\rho &=& \frac{\rho_{0}}{1+\exp([r-R^{'}]/d)} \\
R^{'} &=& R \left[ 1+ \beta_{2}Y_{2}^{0}(\theta)+ \beta_{4}Y_{4}^{0}(\theta) \right]
\end{eqnarray}
where $\rho_{0}$ is the normal nuclear density, $R$ is the radius of the nucleus, $Y_{l}^{m}(\theta)$ denotes the spherical harmonics, $\theta$ is the polar angle and $d$ is the surface diffuseness parameter. For the uranium nucleus $R = 6.81$~fm, $d=0.55$~fm and the surface deformation parameters $\beta_{2} = 0.28$ and $\beta_{4} = 0.093$ are taken from Ref.~\cite{Haque:2011aa}.
Various other published results having studies performed using different Woods-Saxon profiles can be found in Refs.~\cite{Adamczyk:2015obl, Shou:2014eya, PHENIX:2015rkp, PHENIX:2015tbb}.
In Ref.~\cite{Shou:2014eya}  optimization of these parameters was done and the optimized parameters obtained were $R = 6.86$~fm, $d = 0.42$~fm, $\beta_2 = 0.265$, and $\beta_4 = 0$. It was found that for a given $R$, $d$ and $\beta_2$ are anti-correlated. The tuning of parameter $\beta_4$ did not affect the results. Using the new parameters, the maximum $dN/d\eta$ value changed only slightly. In Ref.~\cite{PHENIX:2015rkp}, it was found that there was negligible change in the $N_{\rm{part}}$ values with respect to the old parameters. $N_{\rm{coll}}$ values changed little but were within the uncertainties. Hence, the different sets of parameters would lead to similar results on the particle production.
We have used three different calculations of AMPT calculated data~\cite{Nandi:2019ztz}, obtained from three different partonic cross sections, for which the initial parameter settings are taken from Refs.~\cite{Xu:2011fe, Ma:2016fve} and are shown in Table \ref{tab:lsff}.
Here, $a$ and $b$ are the parameters in the Lund string fragmentation function~\cite{Lin:2000cx} : $f(z) \propto z^{-1} (1-z)^{1} \exp(-b m_{\perp}^{2}/z)$, where $z$ is the light-cone momentum of the produced hadron with transverse mass $m_{\perp}$ with respect to that of the fragmenting string.
Further, based on the Schwinger mechanism for particle production in a strong field, the production probability is proportional to 
$\exp (-\pi m_\perp^2/\kappa)$, where $\kappa$ is the string tension, i.e., energy in a unit length of string, and $\kappa\propto [b(2+a)]^{-1}$.

The parton scattering cross section is $\sigma \approx 9 \pi \alpha_{s}^{2}/(2 \mu^{2})$, where $\alpha_{s}$ is the QCD coupling constant and $\mu$ is the gluon screening mass in QGP.
The three calculated data sets are denoted as AMPTv1 1.5 mb, AMPTv1 10 mb, and AMPTv2 3 mb here and in the following three sections. The statistics analyzed here in the AMPTv1 1.5 mb, AMPTv2 3 3mb set is about 6 M and in the AMPTv1 10 mb set is about 0.4 M.

\begin{table}
  \centering
	\caption{Parameter values in Lund string fragmentation and parton scattering cross section for the three sets of AMPT data used in this paper.}
	\label{tab:lsff}\vspace{0.1in}
	\begin{tabular}{lclclcl}
	\hline	
    \multirow{3}{*}{ Parameter}  &
    \multicolumn{6}{c}{ Cross section $\sigma$} \\ \cline{3-7} 
          && 1.5~mb && 3~mb && 10~mb \\ 
	\hline
	$a$ && 0.5 && 0.55 && 2.2 \\
	$b$~(GeV$^{-2}$) && 0.9 && 0.15 && 0.5 \\
	$\alpha_{s}$ && 0.33 && 0.33 && 0.47 \\
	$\mu$~(fm$^{-1}$) && 3.2 && 2.265 &&  1.8\\
	\hline
	\end{tabular}
\end{table}

\begin{figure*}[!tp]
\centering
\resizebox{1.0\textwidth}{!}{%
  \includegraphics{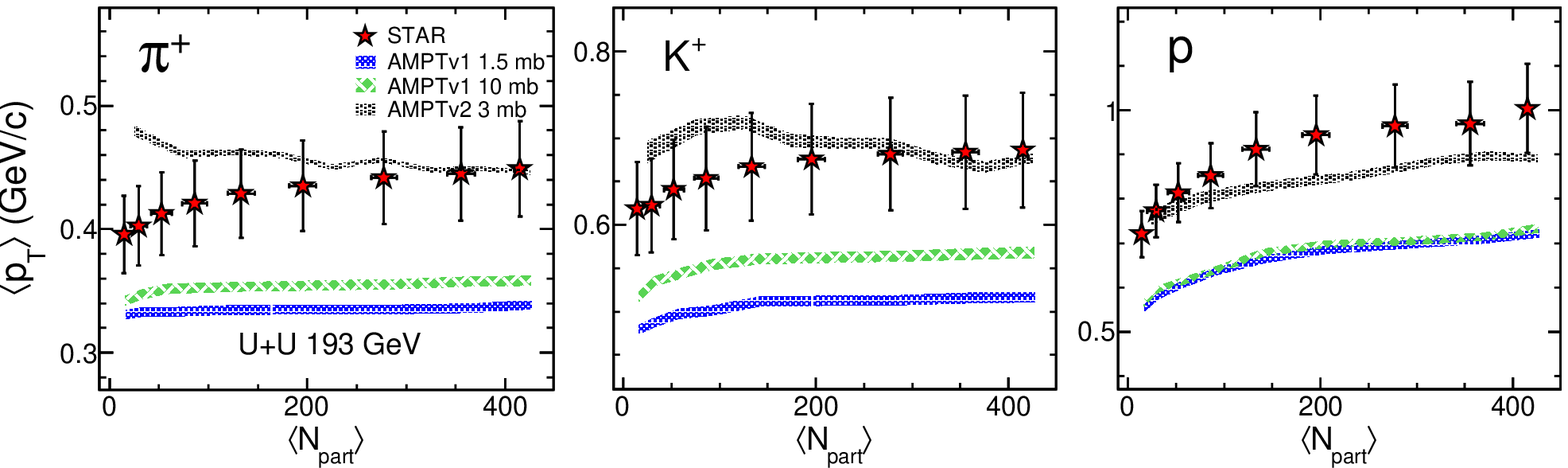}
}
\caption{$\langle p_{\mathrm T}\rangle$ of $\pi^{+}$, $K^{+}$ and $p$ at midrapidity ($|y| < 0.1$) as a function of $\langle N_{\text{part}}\rangle $ obtained for the three sets of AMPT calculated data (AMPTv1 1.5~mb, AMPTv2 3~mb and AMPTv1 10~mb) are shown along with the STAR results obtained in U+U collisions at \snn = 193 GeV. The width of the bands corresponds to the statistical uncertainty associated with the calculation using the AMPT model.}
\label{fig:ampt-mean-pT}
\end{figure*}

\subsubsection{\mdseries{\textit{Mean ${p_{\mathrm T}}$ Comparison}}}

The  $\langle p_{\mathrm T} \rangle$ of $\pi^{+}$, $K^{+}$ and $p$ as a function of $\langle N_{\text{part}}\rangle $ for the three sets of AMPT calculated data are compared with STAR results from U+U collisions at \snn = 193 GeV in Fig.~\ref{fig:ampt-mean-pT}.
The  $\langle p_{\mathrm T} \rangle$ values of all particles are underestimated by AMPTv1 1.5~mb and AMPTv1 10~mb models. The AMPTv2 3~mb model describes the data from mid-central to central collisions for $\pi^+$. For $K^+$ and $p$, the AMPTv2 3~mb model describes the data within uncertainties for all centralities.


\begin{figure*}[!tp]
\centering
\resizebox{1.0\textwidth}{!}{%
\includegraphics{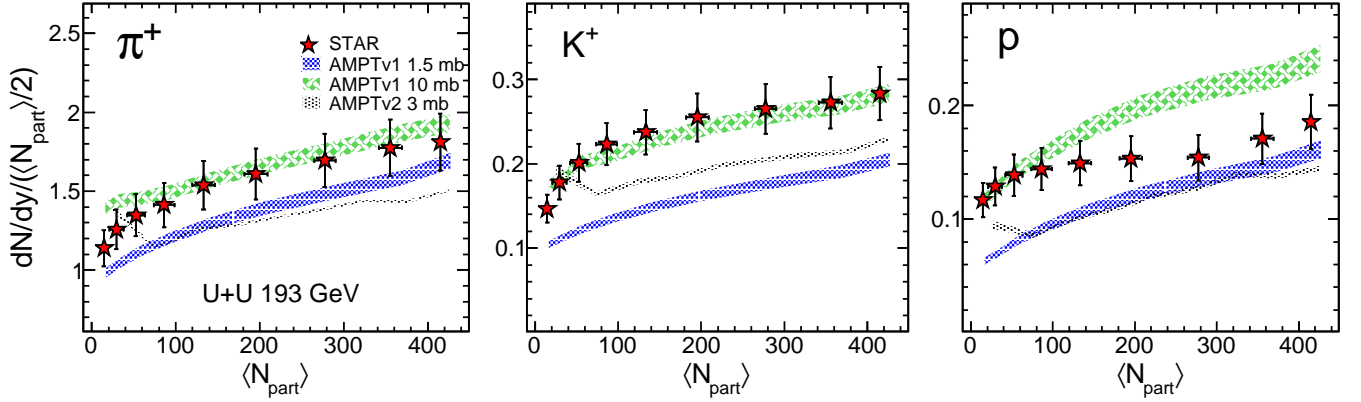}
}
\caption{$dN/dy$ for $\pi^{+}$, $K^{+}$ and $p$ scaled by $\langle N_{\text{part}}\rangle /2$ at midrapidity ($|y| < 0.1$) as a function of $\langle N_{\text{part}}\rangle $ obtained for three calculations of AMPT data (AMPTv1 1.5~mb, AMPTv2 3~mb, and AMPTv1 10~mb) are shown along with the STAR results obtained in U+U collisions at \snn = 193 GeV. The width of the bands corresponds to the statistical uncertainty associated with the calculation using the AMPT model.}
\label{fig:ampt-dndy}
\end{figure*}

\subsubsection{\mdseries{\textit{${dN/dy}$ Comparison}}}
The yields, $dN/dy$ scaled by $\langle N_{\text{part}}\rangle/2 $ for the three calculations of AMPT data are plotted as a function of $\langle N_{\text{part}}\rangle $ as shown in Fig.~\ref{fig:ampt-dndy}.
For comparison, the measured values obtained from U+U collisions at \snn =193 GeV are also shown.   For pions and kaons, the AMPTv1 10 mb case shows decidedly better agreement with the measured values.
For the protons it is seen that agreement between the AMPT calculated values and the measurements depends on the centrality of the collisions with the AMPTv1 10 mb values agreeing better toward peripheral collisions while the AMPTv1 1.5 mb values agree better toward more central collisions.
On the other hand, in case of Au+Au collisions at \snn = 200~GeV, the calculations from the AMPT 1.5~mb case show a better agreement than the AMPT 10~mb case with the charged particle multiplicity density at midrapidity as in Ref.~\cite{Xu:2011fe}.

\begin{figure*}[!tp]
\centering
\resizebox{1.0\textwidth}{!}{%
    \includegraphics{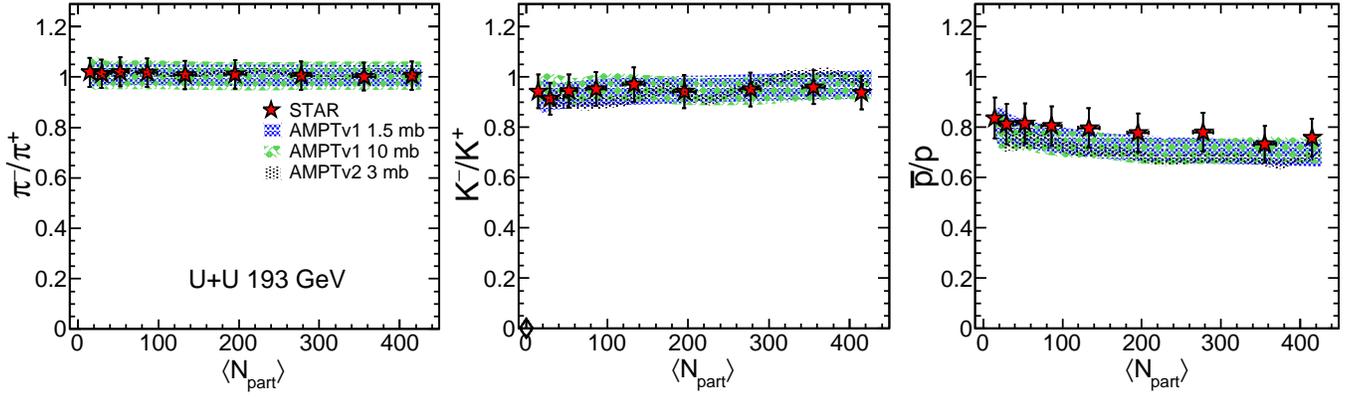}
}
\caption{$\pi^{-}/ \pi^{+}$, $K^{-} / K^{+}$ and $\bar{p}/p$ particle yield ratios at midrapidity ($|y| < 0.1$) as a function of $\langle N_{\text{part}}\rangle $ obtained for three calculations of AMPT data (AMPTv1 1.5~mb, AMPTv2 3~mb, and AMPTv1 10~mb) are shown along with the STAR results obtained in U+U collisions at \snn = 193 GeV. The width of the bands corresponds to the statistical uncertainty associated with the calculation using the AMPT model.}
\label{fig:ampt-like-ratios}
\end{figure*}

\begin{figure*}[!tp]
\centering
\resizebox{0.7\textwidth}{!}{%
    \includegraphics{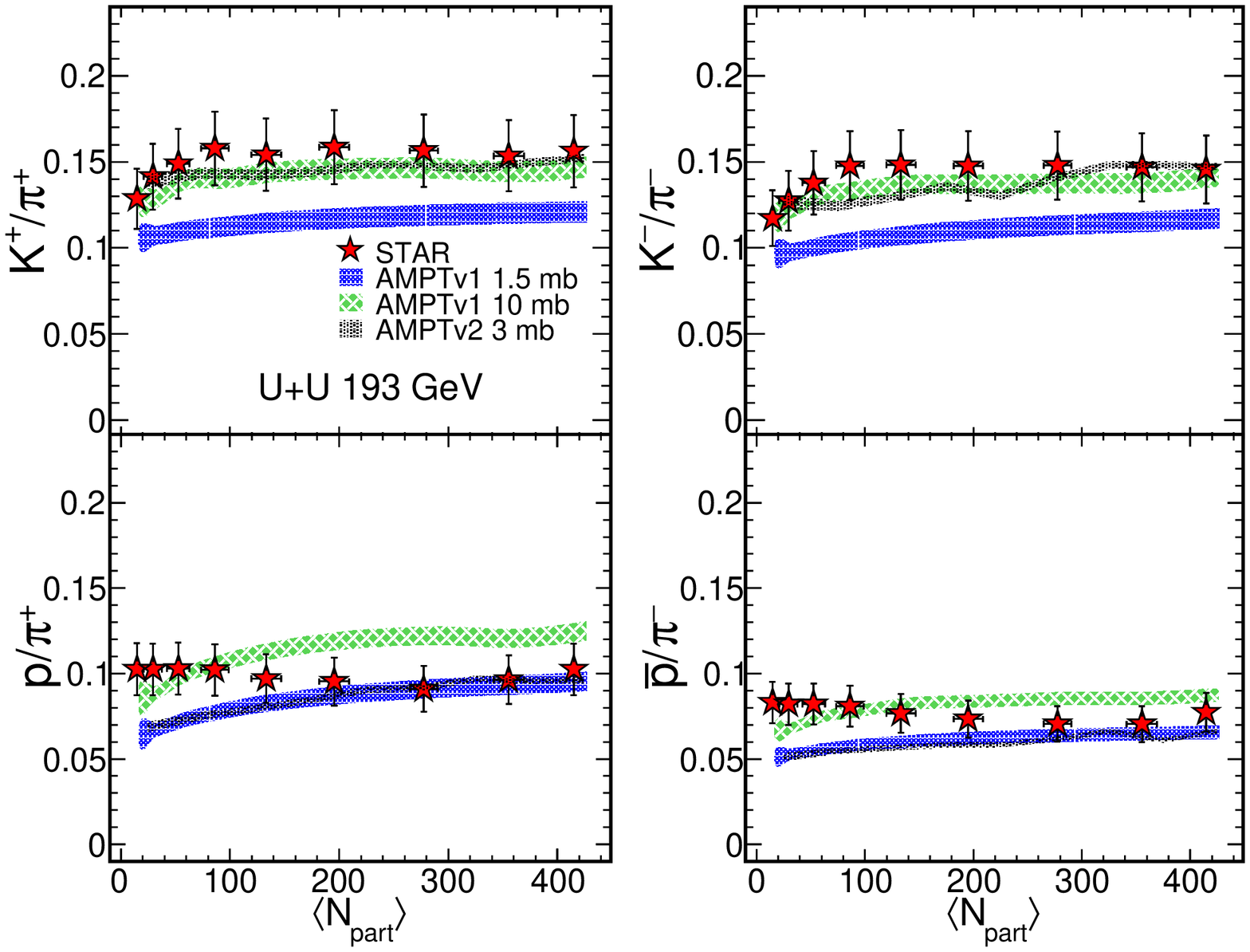}
}
\caption{$K^{+}/ \pi^{+}$, $K^{-} / \pi^{-}$, $p/ \pi^{+}$ and $\bar{p}/ \pi^{-}$ particle yield ratios at midrapidity ($|y| < 0.1$) as a function of $\langle N_{\text{part}}\rangle $ obtained for three calculations of AMPT data (AMPTv1 1.5~mb, AMPTv2 3~mb, and AMPTv1 10~mb) are shown along with the STAR results obtained in U+U collisions at \snn = 193 GeV. The width of the bands corresponds to the statistical uncertainty associated with the calculation using the AMPT model.}
\label{fig:ampt-unlike-ratios}
\end{figure*}

\subsubsection{\mdseries{\textit{Particle Ratios Comparison}}}
The various anti-particle to particle ratios as a function of $\langle N_{\text{part}}\rangle $ obtained from measured data are compared to these ratios that were calculated in the three AMPT data calculations as shown in Fig ~\ref{fig:ampt-like-ratios}. It is observed that these ratios have similar values in all the AMPT cases and agree with STAR data across all centrality classes.

Ratios of unlike particles obtained from the three AMPT data calculations are compared with the STAR results and presented in Fig.~\ref{fig:ampt-unlike-ratios}.  The $K^{+}/ \pi^{+}$ and $K^{-} / \pi^{-}$ ratios for the AMPTv1 10 mb and AMPTv2 3mb cases are in good agreement with the STAR data within errors while the AMPT 1.5 mb case gives values significantly below the STAR data.
For the $p/ \pi^{+}$, $\bar{p}/ \pi^{-}$ cases neither AMPT data set agrees with the data over all centrality classes.  The AMPTv1 10 mb data set agrees better for peripheral collisions while the AMPTv1 1.5 mb and AMPTv2 3mb data set agrees better toward central collisions.


\section{SUMMARY}
Various basic bulk observables regarding identified particle production in U+U collisions at \snn = 193~GeV have been presented. The transverse momentum spectra of $\pi^{\pm}$, $K^{\pm}$ and $p(\bar{p})$ in midrapidity ($|y| < 0.1$) have been measured for nine centrality classes : 0-5\%, 5-10\%, 10-20\%, 20-30\%, 30-40\%, 40-50\%, 50-60\%, 60-70\% and 70-80\%. Other extracted observables from \pt spectra such as average transverse momentum ($\langle p_{\mathrm T} \rangle$), particle yields ($dN/dy$), particle ratios, and kinetic freeze-out properties are also presented as functions of collision centrality or $\langle N_{\text{part}}\rangle$. These observables are compared with the corresponding results from Au+Au collisions at  \snn = 200~GeV and  the AMPT model modified to incorporate the deformation of uranium nucleus.

The mean $\langle p_{\mathrm T} \rangle$ values for $\pi$, $K$, and $p$ increase from peripheral to central collisions in U+U collisions at \snn = 193~GeV. This is an indication of increasing radial flow effects in more central collisions. The increase in $\langle p_{\mathrm T}\rangle$ from $\pi$ to $K$ and to $p$ indicates that the radial flow effect increases with particle mass. 

The integrated particle yields $dN/dy$ in midrapidity $|y| < 0.1$ of $\pi^{\pm}$, $K^{\pm}$, and $p$ do not scale with $\langle N_{\text{part}}\rangle$, but rather they slowly increase from peripheral to central collisions in U+U collisions at \snn = 193~GeV. This indicates that, at this energy, the particle production mechanism has contributions from hard processes involving nucleon-nucleon binary collisions. In contrast, $\bar{p}$ shows no such dependence with centrality of the collision.

Anti-particle to particle yield ratios are close to unity in U+U collisions at \snn = 193~GeV, indicating pair production is the dominant mechanism of particle production at maximum RHIC energy. Their values are also close to those obtained from Au+Au collisions at \snn = 200~GeV. The $K^{+}/ \pi^{+}$ and $K^{-}/ \pi^{-}$ yield ratio increases from peripheral to mid-central collision after which they saturate up to central collisions. The $p/\pi^{+}$ and $\bar{p}/\pi^{-}$ ratios do not show any significant centrality dependence in U+U collisions at \snn = 193~GeV.

The kinetic freeze-out parameters are obtained from the simultaneous Blast-wave fit to $\pi^{\pm}$, $K^{\pm}$, and $p(\bar{p})$ \pt spectra in U+U collisions at \snn = 193~GeV.
The kinetic freeze-out temperature $T_{k}$ decreases from peripheral to central collisions while 
the average flow velocity $\left< \beta \right>$ increases from peripheral to central collisions. This indicates a 
large radial flow effects for central collisions and a short-lived fireball in peripheral collisions.
The extracted $T_{k}$ and $\langle \beta \rangle$ values for similar $\langle N_{\text{part}}\rangle$ values are consistent between Au+Au collisions at \snn = 200~GeV and U+U collisions at \snn = 193~GeV.

A detailed comparison of integrated particle yields, mean transverse momentum, and particle yield ratios from STAR results in U+U collisions at \snn = 193~GeV is carried out with three cases of AMPT string melting corresponding to parton-parton interaction cross sections of 1.5, 3, and 10~mb that was also modified to incorporate the deformation in uranium nucleus.
$\langle p_{\mathrm T} \rangle$ obtained from the AMPTv1 1.5 mb and AMPTv1 10 mb have lower values than STAR results for all centrality classes while AMPTv2 3 mb could describe pions from mid-central to central collisions and kaons and protons for all centralities.
The $dN/dy$ from the AMPTv1 10~mb data are closer to the STAR results for pions and kaons than the AMPTv1 1.5~mb and AMPTv2 3 mb data. However, for protons, $dN/dy$ in central collisions are better explained by the AMPTv1 1.5~mb data but for peripheral collisions there is better agreement with the AMPTv1 10~mb data.
Anti-particle to particle yield ratios are successfully explained by all of the AMPT models implying that they are insensitive to these changes in the parton-parton cross sections.
The $K/\pi$ yield ratios from the AMPTv1 10 mb and AMPTv2 3 mb cases agree with the measurements for all centrality classes, whereas the results from the AMPTv1 1.5 mb case underpredict the STAR results.
In case of $p/\pi$ ratios, AMPTv1 1.5~mb and AMPTv2 3 mb cases more closely explains the STAR results toward central collisions while the AMPTv1 10~mb case seems to be closer to STAR results for peripheral collisions. It can be observed that the three AMPT model cases do not describe all the observables presented here at all $\langle N_{\text{part}}\rangle$ values consistently. The AMPTv2 3 mb model having smaller $a$ and $b$ values describes the $\langle p_{\mathrm T} \rangle$ of data better than others for the $\pi^{\pm}$, $K^{\pm}$, and $p(\bar{p})$. The AMPTv1 10~mb model having largest cross section, describes the $dN/dy$ of $\pi^{\pm}$, $K^{\pm}$, and $p(\bar{p})$ better than the other cases. The antiparticle to particle ratios are described by all the three model cases.
These findings call for further investigation into the AMPT model in terms of the particle production mechanism and the collision dynamics.

We have compared the results from U+U collisions to those from Au+Au collisions  at similar center-of-mass energies, but the colliding nuclei have different geometrical shapes and hence different initial collision conditions. The presented physical observables: $dN/dy$, $\langle p_{\mathrm T}\rangle$ , particle ratios, $T_{k}$ and $\langle \beta\rangle$ are consistent between U+U and Au+Au for a similar range of $\langle N_{\text{part}}\rangle$ values.
This might suggest that, when all the different initial state orientations of the colliding uranium nuclei are combined in the analysis, the resulting final state values approximate what is observed from colliding spherically symmetric nuclei.
In other words, these observables are governed by the $\langle N_{\text{part}}\rangle$ for this analysis.
In future experiments involving colliding uranium nuclei, it would be interesting to measure the different final state observables for identified orientations of the colliding ellipsoidal uranium nuclei.  It has been suggested that this might be done for the most central collisions where the different collision configurations can be selected by a data driven method using selections based on the  probability distribution of the charged particle multiplicity and the event anisotropy for U+U collisions~\cite{Goldschmidt:2015kpa}. In addition, as discussed in Ref.~\cite{Bairathi:2015uba}, the spectator neutron measurements can also be used to select certain initial configurations.

\section{ACKNOWLEDGEMENTS}

We thank Jun Xu and Zi-Wei Lin for discussions related to the AMPT results in the paper.
We thank the RHIC Operations Group and RCF at BNL, the NERSC Center at LBNL, and the Open Science Grid consortium for providing resources and support.  This work was supported in part by the Office of Nuclear Physics within the U.S. DOE Office of Science, the U.S. National Science Foundation, National Natural Science Foundation of China, Chinese Academy of Science, the Ministry of Science and Technology of China and the Chinese Ministry of Education, the Higher Education Sprout Project by Ministry of Education at NCKU, the National Research Foundation of Korea, Czech Science Foundation and Ministry of Education, Youth and Sports of the Czech Republic, Hungarian National Research, Development and Innovation Office, New National Excellency Programme of the Hungarian Ministry of Human Capacities, Department of Atomic Energy and Department of Science and Technology of the Government of India, the National Science Centre and WUT ID-UB of Poland, the Ministry of Science, Education and Sports of the Republic of Croatia, German Bundesministerium f\"ur Bildung, Wissenschaft, Forschung and Technologie (BMBF), Helmholtz Association, Ministry of Education, Culture, Sports, Science, and Technology (MEXT) and Japan Society for the Promotion of Science (JSPS).

\end{document}